\documentclass[onecolumn,11pt]{article}
\usepackage{amsmath}
\usepackage{amsfonts}
\usepackage{amssymb}
\usepackage[usenames,dvipsnames]{pstricks}
\usepackage{epsfig}
\newcommand{\braket}[2]{\langle#1|#2\rangle}
\newcommand{\ket}[1]{|#1\rangle}
\newcommand{\boxit}[1]{\vskip6pt\vbox{\hrule\hbox{\vrule\kern8pt\vbox{\kern8pt\vbox{\hsize344pt\noindent\strut#1\strut}\kern8pt}\kern8pt\vrule}\hrule}}
\title{\textbf{Manifesting the Quantum World}}
\author{Ulrich Mohrhoff\\
\textit{\small Sri Aurobindo International Centre of Education}\\ 
\textit{\small Pondicherry 605002 India}\\
\ttfamily{\small ujm@auromail.net}}
\date{}
\normalsize
\begin{document}
\maketitle

\begin{abstract}
In resisting attempts to explain the unity of a whole in terms of a multiplicity of interacting parts, quantum mechanics calls for an explanatory concept that proceeds in the opposite direction: from unity to multiplicity. Being part of the Scientific Image of the world, the theory concerns the process by which (the physical aspect of) what Sellars called the Manifest Image of the world comes into being. This process consists in the progressive differentiation of an intrinsically undifferentiated entity. By entering into reflexive spatial relations, this entity gives rise to (i)~what looks like a multiplicity of relata if the reflexive quality of the relations is not taken into account, and (ii)~what looks like a substantial expanse if the spatial quality of the relations is reified. If there is a distinctly quantum domain, it is a non-spatial and non-temporal dimension across which the transition from the unity of this entity to the multiplicity of the world takes place. Instead of being constituents of the physical world, subatomic particles, atoms, and molecules are instrumental in its manifestation. These conclusions are based on the following interpretive principle and its more direct consequences: whenever the calculation of probabilities calls for the addition of amplitudes, the distinctions we make between the alternatives lack objective reality. Applied to alternatives involving distinctions between regions of space, this principle implies that, owing to the indefiniteness of positions, the spatiotemporal differentiation of the physical world is incomplete: the existence of a real-valued spatiotemporal background is an unrealistic idealization. This guarantees the existence of observables whose values are real \emph{per se}, as against ``real by virtue of being indicated by the values of observables that are real \emph{per se}.'' Applied to alternatives involving distinctions between things, it implies that, intrinsically, all fundamental particles are numerically identical and thus identifiable with the aforementioned undifferentiated entity.

\end{abstract}

\section{Introduction}
\label{intro}
It seems safe to say that there is a mismatch between what quantum mechanics is trying to tell us about the physical world and how we are programmed---arguably by our very neurobiology---to understand the physical world. The nub of the matter appears to be that while we seek to model physical reality ``from the bottom up'' (i.e., on the basis of some ultimate multiplicity, whether it be a multitude of ultimate building blocks or the multiplicity of points or events in an intrinsically differentiated space or spacetime), what the theory is trying to tell us is that reality is structured ``from the top down.'' In resisting attempts to explain the unity of a whole in terms of a multiplicity of interacting parts, quantum mechanics calls for an explanatory concept that proceeds in the opposite direction, from the unity of an intrinsically undifferentiated entity~$\cal E$ to the multiplicity of the macroworld, by means of an atemporal process of differentiation.

Where I see almost all the other interpretive efforts at an impasse is that none of them seriously asks, ``Why do we have this theory in the first place?'' What I propose in this paper is that we have this theory because it concerns the manifestation of the macroworld. What I mean by the ``macroworld'' is the totality of measurement-independent properties---properties that are real \emph{per se}, as against ``real by virtue of being indicated by values of observables that are real \emph{per se}.'' And what I mean by its ``manifestation'' is its emergence from $\cal E$. By entering into reflexive spatial relations, this intrinsically undifferentiated entity gives rise to (i)~what looks like a multiplicity of relata if the reflexive quality of the relations is not taken into account, and (ii)~what looks like a substantial expanse if the spatial quality of the relations is reified. 

Today it is widely assumed that the classical domain postulated by Bohr should be understood as emergent, and that quantum mechanics ought to explain its emergence. Efforts to understand ``the quantum origins of the classical''~\cite{Zurek2003}, ``the appearance of a classical world in quantum theory''~\cite{Joosetal2003}, or ``the quantum-to-classical transition''~\cite{Schlosshauer07} capitalize on decoherence. Unsurprisingly this approach is not without its critics, for as a purely quantum-mechanical phenomenon, confined to the unitary propagation of correlations, decoherence has no bearing on the existence of the correlata. Unitary dynamics cannot be expected to account for the existence of a domain in which measurements have outcomes.

What is proposed here is that the macroworld emerges not from a quantum domain but from a single entity that transcends categorization. If there is a distinctly quantum domain, it is not a domain of constituent objects but a non-spatial and non-temporal dimension across which the transition from unity to multiplicity takes place. While atoms and subatomic particles are instrumental in this transition, their instrumentality cannot be understood in compositional terms. Ultimately there is but one constituent, to wit,~$\cal E$.

Before I can attempt to substantiate these claims, some housecleaning is in order. It will come as no surprise that Sect.~\ref{sec:mmp} is devoted to the measurement problem. On the face of it, the problem is that
\begin{quote}
according to the unitary development of a quantum dynamics alone, \emph{nothing} does happen in the world: no \emph{click} in a particle detector, no definite measurement outcome, no particle track in a Wilson chamber, no interference pattern at a scintillation screen, and no observable effects of an atom in a Paul trap.~\cite[p.~325, original emphasis]{Falk}
\end{quote}
Either the unitary dynamics is not all that takes place or it is not at all what takes place. Current discussions of the problem~\cite{Mittelstaedt,BLM} assume a tripartite measurement process involving a continuous dynamical process called ``premeasurement.'' A dynamical process continuous in time implies the existence of an intrinsically and completely differentiated spatiotemporal background. Because such a background cannot be considered objective, as will be shown, no unitary dynamics takes place, and the measurement problem in its unsolvable form---the so-called objectification problem---evaporates. The following problem, however, does not. The theory's irreducible empirical core is a probability calculus. This presupposes the events to which, and on the basis of which, it serves to assign probabilities. While, therefore, it cannot be expected to account for their occurrence, it must obviously be consistent with it. The real measurement problem consists in identifying observables whose values are real \emph{per se}, and in explaining how they are distinct from observables that have (definite) values only if (and when) their values are indicated by macroscopic devices, events, or states of affairs.

To be able to address this problem, we need to know why quantum mechanics requires us to use two distinct Lorentz-invariant calculational rules, and we need to pinpoint the essential difference between their respective conditions of application. Why do we have to add amplitudes, rather than probabilities, whenever the conditions stipulated by the second rule are met? This question is answered in the form of a new interpretive principle, which is introduced and discussed in Sect.~\ref{sec:YAIP}.

Applied (in Sect.~\ref{sec:tse}) to alternatives involving distinctions between regions of space, this interpretive principle leads to the conclusion that space cannot be something intrinsically differentiated, something that has parts. What, then, furnishes space with its so-called parts? The short answer is: macroscopic detectors. Suitably amplified, it leads to the conclusion that, owing to the indefiniteness of positions, the spatiotemporal differentiation of the physical world is incomplete---it does not go ``all the way down.'' The existence of a real-valued spatiotemporal background is an unrealistic idealization. This is what makes it possible to identify (in Sect.~\ref{sec:macroscopic}) the observables whose values are real \emph{per se}. They are the positions of macroscopic objects---no surprise there, except that the term ``macroscopic object'' is given a rigorous definition.

Applied (in Sect.~\ref{sec:Identity}) to alternatives involving distinctions between things, the same interpretive principle implies the nonexistence of both diachronic and synchronic individuators. Nothing therefore stands in the way of the claim
that ``identical particles'' are identical in the strong sense of numerical identity, nor of the claim that \emph{all} fundamental particles---not only the indistinguishable ones---are identically the same entity~$\cal E$. The validity of these claims is reinforced in Sect.~\ref{sec:particles}.

Section \ref{sec:lp} addresses theorems to the effect that there is no such thing as a localizable particle~\cite{Hf74,Hf98,Hf2001,Malament,CH}, which has led to the conclusion that particle talk is ``strictly fictional''~\cite{CH}. These theorems are valid if localizability is defined relative to the intrinsically differentiated spatiotemporal manifold presupposed by relativistic quantum field theory, but precisely for this reason they are also irrelevant. Observables that are local relative to this manifold cannot be measured. Real-world detectors monitor regions that are defined relative to the objective system of macroscopic positions, rather than relative to this non-objective manifold. 

Section \ref{sec:Shapes} explains why the idea that a fundamental particle is a literally pointlike object is not only unwarranted on both theoretical and experimental grounds but also inconsistent with the incompleteness of the world's spatial differentiation. The forms of composite objects resolve themselves into spatial relations between \emph{formless} entities. And since these entities are numerically identical, the shapes of things resolve themselves into \emph{reflexive} relations---relations between $\cal E$ and itself. This conclusion paves the way for the principal affirmation of the present paper in Sects.~\ref{sec:lc} and \ref{sec:Manifestation}, according to which quantum mechanics presents us with a new kind of causality. This causality, which underlies the atemporal process of manifestation, is also the only kind that is applicable to the distinctly quantum domain, inasmuch as the temporal concept of causation, which links states or events across time or spacetime, has meaningful application only within the macroworld.

In the final section I discuss how far the central idea of the present paper conforms to what is considered by many to be the most defensible form of scientific realism, to wit, structural realism, particularly the respective versions of ontic structural realism  (OSR) propounded by Ladyman and Ross~\cite{LadyRoss} and by Esfeld and Lam~\cite{EsfeldLam,LamEsfeld}. The weakness of OSR lies in its inability to explain how structure is realized in the physical world. To distinguish between a structure that is physical in some sense from a structure that is purely mathematical, one has to go beyond OSR. What OSR is lacking is the concept of a substance that manifests structure by entering into reflexive relations---both the indefinite relations which are instrumental in the manifestation of the macroworld, and the resulting definite relations which constitute the macroworld.

\section{The measurement problem}
\label{sec:mmp}
The irreducible empirical core of quantum mechanics is a probability calculus. Both the events to which and the data on the basis of which it assigns probabilities are measurement outcomes. That much is common to all formulations of quantum mechanics---Heisenberg's matrix formulation (which is particularly useful in solving harmonic oscillator and angular momentum problems), Schr\"odinger's wave-function formulation (which shifts the focus from observables to states), Feynman's path-integral formulation (which shifts the focus from states to transition probabilities), the density-matrix formulation (which can treat mixed states with ease), Wigner's phase-space formulation (which is particularly useful in considering the classical limit), to name but a few~\cite{Styeretal}.

Though a distinction has to be made between formulations and interpretations of quantum mechanics, the choice of a formulation cannot but bias the range of available interpretations. Current literature on the quantum measurement problem~\cite{Mittelstaedt,BLM}, which still follows the first rigorous formulations of the problem in the monographs of von Neumann~\cite{vonNeumann} and Pauli~\cite{Pauli1933}, is biased toward the wave-function formulation, which according to Styer and coauthors~\cite{Styeretal} ``leaves the conceptual misimpression that [the] wavefunction is a physical entity rather than a mathematical tool.'' Certainly a major factor contributing to this bias is the manner in which quantum mechanics is taught. While a junior-level classical mechanics course devotes a considerable amount of time to different formulations of classical mechanics---Newtonian, Lagrangian, Hamiltonian, least action, etc.---even graduate-level courses emphasize the wave-function formulation almost to the exclusion of all variants.

Current discussions of the measurement problem accordingly proceed on the assumption that there is a measurement process, and that this takes place in three steps: the system or state preparation, a continuous dynamical process called ``premeasurement,'' and the pointer reading or objectification. A dynamical process continuous in time implies the existence of an intrinsically and completely differentiated spatiotemporal background.%
		\footnote{Here Lorentz invariance is assumed to the extent that temporal differentiation and spatial differentiation are mutually implied.}
In this paper I shall widen the range of interpretive options by not taking for granted the existence of such a background. 

Any attempt to go beyond the theory's irreducible empirical core calls for at least one interpretive principle. (Most interpretations incorporate several such principles.) One all but universally accepted principle concerns the ontological status of the coordinate points and instants on which the wave function $\psi(x,t)\equiv\braket x{\psi(t)}$ functionally depends. It is generally taken for granted that these points and instants correspond one-to-one to the elements of an intrinsically and completely differentiated spatiotemporal background.%
		\footnote{At least this is the case outside of attempts at formulating a quantum theory of gravity.}
If this interpretive principle is rejected, the wave function's dependence on time cannot be interpreted as the continuous time-dependence of an evolving instantaneous physical state, inasmuch as this presupposes a completely differentiated spatiotemporal background. The parameter $t$ in $\ket{\psi(t)}$ then can only refer to the time of the measurement to the possible outcomes of which the wave function serves to assign probabilities. Nor is it possible to endorse the eigenvalue-eigenstate link, which Dirac~\cite[p.~46--47]{Dirac} formulated as follows:
\begin{quote}
The expression that an observable ``has a particular value'' for a particular state is permissible in quantum mechanics in the special case when a measurement of the observable is certain to lead to the particular value, so that the state is an eigenstate of the observable.
\end{quote}
In the absence of a completely differentiated spatiotemporal background, it cannot be the case that an observable $A$ has the value~$a$ at every instant~$t$ for which $A\ket{\psi(t)}=a\ket{\psi(t)}$ holds. The information provided by this eigenvalue equation is conditional: if $A$ is measured at the time $t$ then  the value $a$ is found to be possessed \emph{at the time} $t$ with probability~1. This also means that probability~1 is not sufficient for ``is'' or ``has.'' For a value to be possessed, a measurement has to be made.

Hence, if we use the wave-function formulation to assign a probability $p$ to a possible outcome $q_2$ of a measurement made at the time $t_2$, on the basis of the actual outcome $q_1$ of a measurement made at the time~$t_1$, we should think of the wave function not as an evolving physical state but as a ``computing machine'' with inputs and outputs: if we plug in $q_1$, $t_1$, $q_2$, and $t_2$, then $p$ pops out. (If the measurement at $t_1$ is not a complete measurement, $p$ will also depend on outcomes of other measurements.) It bears repetition: $t_1$ and $t_2$ refer to the times of the measurements on the basis of which, and to the possible outcomes of which, the wave function serves to assign probabilities. As the inputs $q_1$ and $q_2$ are given by macroscopic ``pointers,'' so the times $t_1$ and $t_2$ are given by macroscopic clocks.%
	\footnote{The following question has been asked by a reviewer (to whom I am grateful for many valuable suggestions): If the wave function does not describe \emph{some} kind of (holistic) physical reality, how is it possible for experiments to act on the wave function in preparation procedures? The way I see it, the wave function is not something on which experiments can act. Preparation procedures \emph{define} wave functions, which serve to assign probabilities to the possible outcomes of measurements. On the other hand, nothing stands in the way of positing such a reality, and of thinking of the wave function as a means to describe it \emph{by assigning probabilities to the possible outcomes of measurements}. We may think of a preparation procedure as acting on such a reality, its effect being what the wave function so describes.}

While the objectification problem evaporates if no intrinsically and completely differentiated spatiotemporal background is postulated, the measurement problem does not. Quantum theory's irreducible empirical core presupposes the events to which, and on the basis of which, it serves to assign probabilities. While, therefore, it cannot be expected to account for their occurrence, it has to be consistent with it. For this to be the case, it must be possible to look upon the values of certain observables as real \emph{per se}, as against ``real by virtue of being indicated by values of observables that are real \emph{per se}.'' The real measurement problem consists in identifying observables whose values are real \emph{per se}, and in explaining how they are distinct from observables that have (definite) values only if (and when) their values are indicated by macroscopic devices, events, or states of affairs. To be able to address this problem, however, we first need to know why quantum mechanics requires us to use two distinct Lorentz-invariant calculational rules, and we need to pinpoint the essential difference between their respective conditions of application.

\section{Yet another interpretive principle}
\label{sec:YAIP}
A salient feature of the quantum-mechanical probability calculus is the nonexistence of dispersion-free probability algorithms. A particular instance of this feature is the uncertainty principle. Feynman, letting the particular stand for the general, defines the uncertainty principle in terms of its observable consequences:
\begin{quote}
Any determination of the alternative taken by a process capable of following more than one alternative destroys the interference between alternatives.~\cite[p.~9]{FHS}
\end{quote}
In a previous paper~\cite {Mohrhoff-QMexplained} I formulated Feynman's version in the following way, taking into account that the mere possibility of determining the alternative taken can ``destroy'' the interference between alternatives.
\begin{description}
\item[{Premise A}.]Quantum mechanics provides us with algorithms for assigning probabilities to possible measurement outcomes on the basis of actual outcomes. Probabilities are calculated by summing over alternatives. Alternatives are possible sequences of measurement outcomes. Associated with each alternative is a complex number called ``amplitude.''
\item[{Premise B}.]To calculate the probability of a particular outcome of a measurement~$M_2$, given the actual outcome of a  measurement~$M_1$, choose a sequence of measurements that may be made in the meantime, and apply the appropriate rule.
\item[{Rule A}.]If the intermediate measurements are made (or if it is possible to infer from other measurements what their outcomes would have been if they had been made), first square the absolute values of the amplitudes associated with the alternatives and then add the results.
\item[{Rule B}.]If the intermediate measurements are not made (and if it is not possible to infer from other measurements what their outcomes would have been), first add the amplitudes associated with the alternatives and then square the absolute value of the result.
\end{description}
The need for the parenthetical phrases in Rules~A and~B can be illustrated with the ``quantum eraser'' experiment discussed by Englert, Scully, and Walther~\cite{SEW1991,ESW1994,Mohrhoff-ESW}. As long as the two microwave resonance cavities are separated, the photon makes it possible to obtain which-way information, and this possibility ``destroys'' the interference between the alternatives.%
		\footnote{It is often stated that the photon (or the cavity field) stores (or contains) which-way information, but this is misleading at best. Strictly speaking, only an actual state of affairs can contain information. The photon only makes it possible to obtain that information, by detecting it in either of the cavities. The detection of the photon in either cavity creates what the information conveys, namely the fact that the atom went through the corresponding slit.}
		
It should be noted that Premise A defines alternatives in terms of measurement outcomes. The only referents needed to formulate the laws of quantum mechanics are measurement outcomes and their correlations. However, the events that the theory correlates are measurement outcomes only in the restricted sense that they indicate---make available information about---the values of observables, irrespective of an experimenter's intentions, and irrespective of the presence of an observer. It is also noteworthy that the time at which $M_2$ is made need not be later than the time at which $M_1$ is made. The quantum-mechanical probability calculus allows us to assign probabilities not only to the outcomes of later measurements on the basis of the outcomes of earlier measurements but also vice versa. As agents in a successively experienced world, we are of course more interested in predictions than in postdictions, but this temporal asymmetry is external to the theory's irreducible empirical core. It justifies the notion that quantum states evolve towards the future as little as it justifies the notion that quantum states evolve towards the past.

From the point of view adopted by the standard approach to the measurement problem, Rule~B seems uncontroversial. Superpositions are ``normal,'' and what is normal does not call for explanation. What calls for explanation is the existence of a mixture that admits of an ignorance interpretation. According to the present approach, the uncontroversial rule is Rule~A, inasmuch as it is what classical probability theory leads one to expect. What calls for explanation is why we have to add amplitudes, rather than probabilities, whenever the conditions stipulated by Rule~B are met. To this question I propose the following answer:
\begin{description}
\item[YAIP]Whenever quantum mechanics instructs us to first add the amplitudes associated with alternatives and then square the absolute value of the result, the distinctions we make between the alternatives correspond to nothing in the physical world.
\end{description}
The converse does not hold. Situations in which quantum mechanics instructs us to add probabilities do not automatically warrant that the distinctions we make between the alternatives are objective.%
		\footnote{The experiment discussed by Englert, Scully, and Walter provides a counterexample. As long as the photon is inside the union of the two resonance cavities, the probability of detecting the atom at the screen is given by Rule~A, yet the atom cannot be said to have taken a particular slit. This becomes clear when interference is restored, by (i)~opening the electro-optical shutters that separate the two cavities and (ii)~sorting the detected atoms according as the photosensor situated between the shutters does, or does not, respond.}

In making the theory's irreducible empirical core my starting point, I do not intend to advocate a radical empiricism or a metaphysically sterile instrumentalism; my only reason for doing so is to clear the terrain of unwarranted assumptions.%
	\footnote{Chris Fuchs~\cite[p.~46]{Fuchs} knows all about being accused of instrumentalism: ``Believe me, you've got to stand your ground with these guys when their label guns fly from their holsters! I say this because if one asks `Why the quantum?' in this context, it can only mean that one is being realist about the reasons for one's instrumentalities. In other words, even if quantum theory is purely a theory for apportioning and structuring degrees of belief, the question of `Why the quantum?' is nonetheless a question of what it is about the actual, real, objective character of the world that compels us to use this framework for reasoning rather than another.''}
I~am aware that we cannot conceive of or discuss any subject matter independently of our thoughts or our language. A completely mind-independent reality would be epistemically inaccessible. However, I am not concerned with the question whether \emph{mind}-independent reality can be attributed to the physical world. What I am interested in is how to identify those observables to which \emph{measurement}-independent reality can be attributed. 

While YAIP imposes limits on the extent to which our conceptual distinctions can be considered objective, it is not intended as being merely a statement of our practical or conceptual limitations. To say that our theories are constructions is to state the obvious, but it does not explain why most of our falsifiable constructions turn out to be false. Each time we learn a way things are \emph{not}, we come closer to knowing the way things are; knowledge is most objective precisely when it tells us where we are wrong.

To illustrate this point, the epistemologist von Glasersfeld~\cite{vG} has made use of the difference between a match and a fit. He imagined a skipper who, in the dark of a stormy night, without navigational aids, passes a narrow strait whose contour he does not know. Epistemologically, we are in the skipper's position. If he reaches the open sea without mishap, he has found a course that \emph{fits} the strait; if next time he takes the same course, he will again pass safely. What he has not obtained is a map that \emph{matches} the coastline. To precisely locate at least one point of the coastline, he must come into contact with it---at the risk of wrecking his ship. What YAIP asserts is that whenever quantum mechanics requires the use of Rule~B, we ``wreck our ship'' by attributing objectivity to the distinctions we make between the alternatives.

The realism I am defending is a realism in the moderate and epistemologically justified sense of a fit. It is first of all a realism about the macroworld. It asserts the measurement-independence of the macroscopic observables that will be defined in Sect.~\ref{sec:macroscopic}, without infringing the universal validity of quantum mechanics. It is also a realism about the measured values of microphysical observables. Unlike the minimal interpretation as defined by Mittelstaedt~\cite[pp.~9--11]{Mittelstaedt}, which ``avoids statements about object systems and their properties and instead refers to observed data only'' (i.e., to pointer values), it allows statements about the values of observables---if and when they are measured. Unlike the realistic interpretation as defined by Mittelstaedt~\cite[pp.~12--14]{Mittelstaedt}, it eschews the notion that the measured value of an observable ``pertains actually to the object system after the measurement,'' which amounts to an endorsement of the eigenvalue-eigenstate link. It is, moreover, a realism about a single independently existing entity~$\cal E$, which manifests the macroworld by entering into reflexive spatial relations, as will be spelled out in Sects.~\ref{sec:Identity} through~\ref{sec:Manifestation}.

\section{A two-slit experiment and its implications}
\label{sec:tse}
In this section the interpretive principle formulated in the previous section will be applied to the two-slit experiment with electrons, which according to Feynman~\cite[Sect. 1.1]{Feynman65} ``has in it the heart of quantum mechanics'' and ``is impossible, absolutely impossible, to explain in any classical way.'' 

If Rule B applies, then according to YAIP the distinction we make between $\mathbf{p}_L$~=~``the electron went through the left slit~($L$)'' and $\mathbf{p}_R$ = ``the electron went through the right slit~($R$)'' cannot be considered objective. It is not the case that a given electron goes through either $L$ or~$R$. (Because the build-up of the tell-tale interference pattern has been demonstrated one electron at a time~\cite{Tonomura89}, it is permissible here to refer to a ``given'' electron.) Somehow an electron can go through both slits---as a whole, without being divided into parts that go through different slits. But how? 

Our difficulty in understanding how this is possible reveals something peculiar about how we tend to think about space. We are inclined to think that $L$ and $R$ are intrinsically distinct. But how do they differ? They are cutouts in a slit plate---things that have been removed, things that are no longer there. What difference do they leave behind once they have been removed? The difference between the positions they previously occupied? But positions are properties, and properties exist only if they are possessed. 

How does a physical property come to be possessed? Being the outcome of a measurement is sufficient for a property to be possessed---at the time of the measurement---but is it also necessary? We have seen that it is. Hence $L$ (or~$R$) only exists (as a possessed property) if Rule~A applies, and if the electron is found to have taken the left (or right) slit.

If the distinction we make between the propositions $\mathbf{p}_L$ and $\mathbf{p}_R$ cannot be considered objective, then the distinction between $L$ and $R$ cannot be real \emph{per se}. The electron's position at the (indeterminate) time when it passes the slit plate is the union $L\cup R$. If this were objectively divided into distinct regions $L$ and~$R$, the electron's position at that time would be affected by the distinctness of $L$ and~$R$; it would be divided by it. Yet a position is not something that can be divided.%
	\footnote{What is at issue here is the divisibility of a position. Interference fringes have been observed using C$_{60}$ molecules and a grating with 50-nm-wide slits and a 100-nm period~\cite{Arndtetal}. We do not picture parts of such a molecule as getting separated by many times 100 nm and then reassembling into a ball less than a nanometer across.}
The indivisibility of the electron's position therefore implies the indivisibility of $L\cup R$.

But if $L\cup R$ is not intrinsically divided, then physical space cannot be intrinsically divided. If at all we think of it as a self-existent or substantial expanse, we must think of it as undifferentiated, without parts. The question then is, what furnishes space with its so-called parts? Or rather: what furnishes the physical world with its spatial parts? The short answer is: detectors. Detectors, and more generally measuring devices, exist not only in man-made laboratories. Any device with a sensitive region $D$ and capable of indicating the presence of something in $D$ qualifies as a detector, and any device capable of providing information about the value of an observable qualifies as a measuring device. An idealized position measurement uses an array of detectors, and if projection-operator-valued probability-measures are used, the detectors' sensitive regions $D_i$ correspond to a partition of some larger region. If an object is found in~$D_i$ (but not in any smaller region inside~$D_i$), the position attributable to it is $D_i$---not any smaller region inside~$D_i$, let alone a sharp position.

By those of its macroscopic properties that define a spatial region~$D$, a detector realizes~$D$, and by realizing a spatial region~$D$, a detector makes it possible to attribute to a microphysical object the property of being in~$D$. That this bears generalization was stressed by Bohr~\cite{Bohr34,Bohr49,Bohr63}: the measurement apparatus is needed not only to indicate the possession of a property (by a system) or a value (by an observable), but also, and in the first place, to make a set of properties or values available for attribution. In the absence of an apparatus that realizes a set of properties~$q_i$, the properties~$q_i$ are not possible attributes, and the distinctions we make between the statements ``system $S$ has the property~$q_i$'' cannot be considered objective. (The present paper goes beyond Bohr in that it does not split the world into a classical and a quantum domain but instead places the domain of macroscopic observables squarely within the quantum domain.)
 
As an illustration, consider the measurement of a component of the spin of a particle of spin-$1/2$. In this case the apparatus serves not only to indicate an outcome---the component's value at the time of measurement---but also, and in the first place, to realize an axis (by means of the gradient of a magnetic field) and to thereby make two possible values available for attribution. It creates possibilities to which probabilities can be assigned. Without an apparatus that defines an axis, the two values do not even exist as possibilities.%
		\footnote{\label{note-hydrogen}Generations of students have been puzzled by the special role that the $z$~axis plays in descriptions of the stationary states of atomic hydrogen. How does the atom chose this particular axis? The answer, of course, is that it doesn't. Quantum-mechanical probability assignments are conditional on preparations. In describing the atom's stationary states we \emph{assume} that the $z$~component of its angular momentum has been measured, along with its energy and its total angular momentum.}

It will be instructive to briefly concern ourselves with the question how we come to believe that positions exist by themselves, without being possessed. After all, we more or less readily agree that red, round, or a smile cannot exist without a red or round object or a smiling face. This is why the Cheshire cat strikes us as funny. Why are positions treated differently? Why don't self-existent positions make us laugh? What comes to mind is that the role position plays in perception is analogous to the role substance plays in conception. 

For Aristotle, a property was anything in the world that can be the predicate of a sentence composed of a subject and a predicate, whereas a substance was something in the world that cannot be predicated of anything else. Substance, so defined, serves two purposes: it betokens independent existence, and it reifies the manner in which a conjunction of predicative sentences with the same subject term bundles predicates. (Property-bundling substances are often referred to as Lockean substances.)

Turning to the neuroscientific data~\cite{Hubel95,Enns04}, one is struck by the abundance of feature maps. A feature map is a layer of the cerebral cortex in which cells map a particular phenomenal variable (such as hue, brightness, shape, viewer-centered depth, motion, or texture) in such a way that adjacent cells generally correspond to adjacent locations in the visual field. In the macaque monkey, as many as 32 distinct feature maps have been identified~\cite{Clark}. Every phenomenal variable has a separate map (and usually not just one but several maps at different levels within the neuro-anatomical hierarchy) except location, which somehow is present in all maps. If there is a green box here and a red ball there, ``green here'' and ``red there'' are signaled by neurons from one feature map, and ``boxy here'' and ``round there'' are signaled by neurons from another feature map. ``Here'' and ``there'' are present in both maps, and this is how we know that green goes with boxy and red goes with round. While it is still far from clear how feature integration is achieved by the brain, it is clear that position is the integrating factor.

Thus while substance serves as the ``conceptual glue'' that binds an object's properties, position serves as the ``perceptual glue'' that binds an object's phenomenal features. Failure to distinguish between perceptual objects and conceptual objects, or between the two types of ``glue,'' therefore appears to be at least partly responsible for the substantivalist conception of space,%
		\footnote{\label{note:feature-integration}It is also what makes us conceive of features present in the same place as features of the same object, and to conceive of features present in different places as features of different objects (or of different parts of the same object), so that we are thoroughly baffled by the ability of an indivisible object to pass simultaneously through different slits.}
which in the second half of the nineteenth century merged with the set-theoretic conception of space. On the latter conception von Weizs\"acker~\cite[p.~130]{vW1980} remarked:
\begin{quote}
The conception of the continuum as potential, which originated with Aristotle, appears to be more suitable for the quantum theoretical way of thinking than is the set-theoretical conception of an actually existing transfinite manifold of ``real numbers,'' or of the spatial points they designate. The ``real number'' is a free creation of the human mind and perhaps not conformable to reality.
\end{quote}
If proof is needed that the set-theoretic conception of space is not conformable to reality, it is the ability of a particle to simultaneously go through more slits than one. If there is a single system $S$ for which the distinction between ``$S$~went through~$L$'' and ``$S$~went through~$R$'' cannot be considered objective, then the distinction we make between $L$ and~$R$ cannot be so considered, let alone all of the distinctions that are implicit in the set-theoretic conception of space. 

The idealized detectors consider in this section do not exist. As a consequence of the uncertainty principle, sharply localized and sharply bounded spatial regions cannot be realized. The kind of position measurement that uses an array of detectors with sharply localized and sharply bounded regions is a heuristic fiction. But if such regions cannot be realized (as the sensitive regions of detectors), then they cannot be attributed (as positions). It follows that the spatial differentiation of the physical world cannot be complete---it cannot go ``all the way down.'' We can conceive of a partition of the physical world into \emph{finite} regions so small that none of them can be attributed (as a position) because none of them is available for attribution.

The same goes for the world's temporal differentiation, and this not only because of the relativistic interdependence of distances and durations. Just as the properties of quantum systems or the values of quantum observables need to be realized---made available for attribution---by macroscopic systems, so the times at which properties or values are possessed need to be realized by macroscopic clocks. And just as it is impossible for macroscopic systems to realize sharp positions, so it is impossible for macroscopic clocks to realize sharp times~\cite{Hilgevoord98}. Therefore neither the spatial nor the temporal differentiation of the physical world goes ``all the way down.''

\section{The macroworld}
\label{sec:macroscopic}
If quantum theory is to accommodate value-indicating events or states of affairs, it must be possible to look upon the values of certain observables as real \emph{per se}, as against ``real by virtue of being indicated by the values of observables that are real \emph{per se}.'' For this, something has to give. Neither can actually possessed positions be sharp, nor can all unsharp positions be merely distributions over possible outcomes of unperformed measurements. There has to be a middle ground; there must be unsharp positions that are actually possessed, rather than being merely probability distributions over unrealized possibilities. And indeed, as I shall argue presently, there is a (non-empty) class of objects whose unsharp positions remain unresolved by measurements.

In a world that is incompletely differentiated spacewise, the next best thing to a sharp trajectory is a trajectory that is so sharp that the bundle of sharp trajectories over which it is statistically distributed is never probed. In other words, the next best thing to an object with a sharp position is an object whose position probability distribution is and remains so narrow that there are no detectors with narrower position probability distributions---detectors that could probe the region over which the object's unsharp position extends. If the spatiotemporal differentiation of the physical world does not go ``all the way down,'' such objects must exist. What shall we call them? They are not ``classical'' because classical objects follow sharp trajectories. They are not ``macroscopic'' in the sense of being so large and/or massive as to behave like classical objects FAPP (for all practical purposes). If I call them ``macroscopic objects,'' it is in the more rigorous sense just spelled out. The \emph{macroworld}, as the term is used in this paper, is the totality macroscopic positions. (``Macroscopic positions'' is short for ``positions of macroscopic objects.'')

What can be deduced from this characterization of macroscopic positions is that the events by which their values are indicated are (diachronically) correlated in ways that are consistent with the laws of motion that quantum mechanics yields in the classical limit. For any given time $t$ the following holds: if every event that indicates a macroscopic position prior to the time $t$ were taken into account, then---given the necessarily finite accuracy of position-indicating events---every event that indicates a macroscopic position at a later time would be consistent with all earlier position-indicating events and the classical laws motion. There is, however, one exception: to permit a macroscopic object---the proverbial pointer---to indicate the value of an observable, its position must be allowed to change unpredictably if and when it serves to indicate a value.

Macroscopic objects thus follow trajectories that are only counterfactually indefinite. Their positions are ``smeared out'' only in relation to an imaginary spatiotemporal background that is more differentiated than the physical world. No value-indicating event reveals the indefiniteness of a macroscopic position (in the only way it could, through a departure from what the classical laws predict). The testable correlations between outcomes of measurements of macroscopic positions are therefore consistent with both the classical \emph{and} the quantum laws. This is what makes it possible to stipulate that macroscopic positions are real \emph{per se}, in a way that is not liable to Bell's~\cite{Bell1990} critique of FAPP solutions to the measurement problem. 

A similar stipulation has recently been proposed by Bub~\cite{BubBW}:
\begin{quote}
The  problem of how to account for the definiteness or determinateness of the part of the universe that records the outcomes of quantum measurements or random quantum events is a consistency problem. The question is whether it is consistent with the quantum dynamics to take some part of the universe, including the registration of quantum events by our macroscopic measuring instruments, as having a definite ``being-thus,'' characterized by definite properties.
\end{quote}
Bub's answer to this question is affirmative:
\begin{quote}
If we take [the preferred observable] $R$ as the decoherence ``pointer'' selected by environmental decoherence, then it follows that the macroworld is always definite because of the nature of the decoherence interaction coupling environmental degrees of freedom to macroworld degrees of freedom (a contingent feature of the quantum dynamics), and it follows from the theorem%
		\footnote{The theorem referred to by Bub is proved in Ref.~\cite{BubClifton}.}
that features of the microworld correlated with $R$ are definite. In other words, decoherence guarantees the continued definiteness or persistent objectivity of the macroworld, if we stipulate that $R$ is the decoherence ``pointer.''
\end{quote}
Bub's answer, however, is liable to the objection~\cite{MohrhoffBW} that taking the pointer selected by environmental decoherence as definite by stipulation amounts to deliberately ignoring the off-diagonal elements of the density operator for the larger system that includes the environment. This may be justifiable FAPP but remains subject to Bell's critique of FAPP solutions to the measurement problem. Bub~\cite{BubBW} has countered this objection, insisting that
\begin{quote}
The argument here is not that decoherence provides a dynamical explanation of how an indefinite quantity becomes definite in a measurement process---Bell \cite{BellCH} has aptly criticized this argument as a `for all practical purposes' (FAPP) solution to the measurement problem.%
		\footnote{In the paper cited, Bell examines a paper by Hepp~\cite{Hepp} whose abstract contains the following statement: ``In several explicitly soluble models, the measurement leads to macroscopically different `pointer positions' and to a rigorous `reduction of the wave packet' with respect to all local observables.''}
\end{quote}
Bell's critique of FAPP solutions~\cite{Bell1990}, however, is not limited to quantum dynamical accounts of the emergence of classical behavior. It applies whenever small quantities are treated as zero on the ground that their consequences are unobservable FAPP, and in particular to any attempt to solve the measurement problem by invoking environment-induced decoherence.

Decoherence-based solutions to the measurement problem are suspect for another reason. As critics of this approach have pointed out, among them Mittelstaedt~\cite[p.~112]{Mittelstaedt} and Busch et al.~\cite[p.~123]{BLM}, the quantum-mechanical coherence of the system composed of apparatus and object system is merely displaced into the degrees of freedom of the environment. The objectification problem reappears as a statement about the system composed of environment, apparatus, and object system. Since the mixture obtained by tracing out the environment does not admit an ignorance interpretation with respect to the pointer basis, environment-induced decoherence does not solve the objectification problem.

Nor is there any need to solve this pseudo-problem. If one starts by assuming that quantum states are evolving physical states, one needs something more than physical to get from possibilities to actualities, and this miracle cannot be wrought by decoherence (nor, for that matter, by anything else). The existence of actualities is not in question. What environment-induced decoherence does is to quantitatively support the conclusion that macroscopic objects (as herein defined) exist, by showing that for macroscopic objects (as defined by various quantitative models) the probability of obtaining evidence of departures from classical behavior is extremely low. This guarantees the abundant existence of macroscopic objects (as herein defined).

The standard representation of the values of observables by means of projection operators is not suited to dealing with detectors whose sensitive regions have unsharp boundaries. The appropriate formalism for dealing with such detectors uses positive-operator-valued (POV) measures~\cite{BGL}. Busch and Shimony~\cite{Bushi,Bush96} (see also Ref.~\cite[Sect. III.6.2]{BLM}) have shown that the objectification problem remains unsolvable if POV measures are used instead of projector-valued ones---unless the pointer observable itself is unsharp:
\begin{quote}
in using the general representation of observables as POV measures, all kinds of inaccuracy have been taken into account---to the extent they are still compatible with the idea of definite pointer values. The remaining potential loophole is furnished by the case of pointer observables which are genuinely unsharp in that they do not allow for pointer value definiteness.~\cite{Bush96}
\end{quote}
That pointer observables are unsharp and pointer values indefinite is a direct consequence of the incomplete spatiotemporal differentiation of the physical world. However, pointers are macroscopic objects, and macroscopic positions are unsharp only in relation to an imaginary spatiotemporal background that is more differentiated than the physical world. This is why we are justified on principled grounds to look on the transition of a pointer from its neutral position to a value-indicating position as the kind of actual event without which the quantum-mechanical probability calculus would lack a domain of application.

In an attempt to make sense of unsharp pointers in the framework that gives rise to the objectification problem, Mittelstaedt~\cite[p.~116, original emphases]{Mittelstaedt} conceives of indefinite pointer values as ``\emph{almost real} properties'' and argues that 
\begin{quote}
\emph{one has to give up the idea of an objective reality.} Indeed, not only microphysical systems but also macroscopic instruments and pointers would be in a state of objective undecidedness that is expressed by the genuine unsharpness of the pointer observables. Even if the degree of objectification is very high in all practical cases the observations will always contain a finite amount of nonobjectivity\dots. The use of genuinely unsharp pointer observables\dots must not be considered as a means to restore objectivity and reality in the physical world.
\end{quote}
Giving up the idea of an objective reality amounts to giving up the idea that the quantum-mechanical probability calculus has a domain of application. What leads to such self-contradictory expressions as ``almost real properties'', a ``degree of objectification'', or an ``amount of nonobjectivity'' is the identification of reality/ objectivity with definiteness. Reality and objectivity do not come in degrees, nor does a pointer position have to be sharp in order to be real. What is required for it to be objective is that it be unsharp only in relation to an imaginary spatiotemporal background that is more differentiated than the physical world.

\section{Particles: identity and individuality}
\label{sec:Identity}
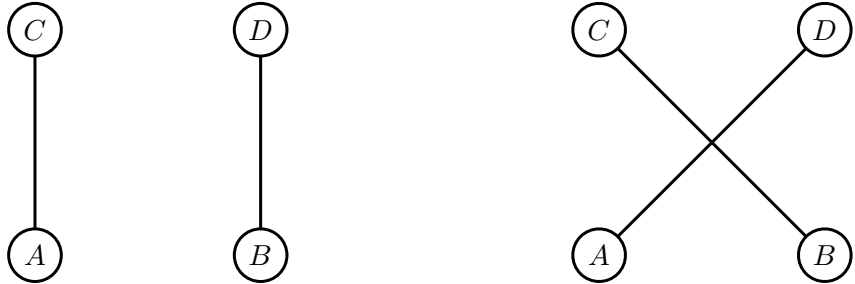
\begin{figure}[t]
\scalebox{1}
{
\begin{pspicture}(-0.55,-2)(10,2)
\psline[linewidth=0.04cm,dotsize=0.66cm 2.0]{o-o}(0.0,1.5)(0.0,-1.5)
\psline[linewidth=0.04cm,dotsize=0.66cm 2.0]{o-o}(3,1.5)(3,-1.5)
\rput(0,1.5){$C$}
\rput(0,-1.5){$A$}
\rput(3,1.5){$D$}
\rput(3,-1.5){$B$}
\psline[linewidth=0.04cm,dotsize=0.66cm 2.0]{o-o}(7.5,1.5)(10.5,-1.5)
\psline[linewidth=0.04cm,dotsize=0.66cm 2.0]{o-o}(10.5,1.5)(7.5,-1.5)
\rput(7.5,1.5){$C$}
\rput(7.5,-1.5){$A$}
\rput(10.5,1.5){$D$}
\rput(10.5,-1.5){$B$}
\end{pspicture}
}
\caption{Two alternatives. If quantum mechanics instructs us to add their amplitudes rather than their probabilities, then the straight lines, which represent transtemporal individuators of some kind, do not correspond to anything in the physical world.}
\label{fig-alt}
\end{figure}
Our next order of business is to apply YAIP to alternatives involving distinctions between things. To this end we consider four non-overlapping regions $A$, $B$, $C$, $D$, realized by the sensitive regions of four macroscopic detectors. Initial measurements (performed at the time~$t_1$) indicate the presence of one particle in $A$ and one particle in~$B$. We wish to calculate the probability $p$ with which subsequently (at the time~$t_2$) one particle is found in $C$ and one in~$D$. There are two alternatives (Fig.~\ref{fig-alt}). Whenever quantum mechanics instructs us to add their amplitudes, i.e., $p=|A_1+A_2|^2$, the distinction we make between the alternatives cannot be considered objective, and the lines connecting the four regions have no counterparts in the physical world. These lines represent diachronic individuators---either persistent individuating properties or individuating substances. If they cannot be considered objective, then diachronic individuators do not exist. 

Absent diachronic individuators, the particles observed at $t_1$ and the particles observed at $t_2$ cannot be ``snapshots'' of things that are distinct \emph{and remain so} as time advances from $t_1$ to $t_2$. Are there synchronic individuators, over and above the different regions containing the particles at $t_1$ and again at $t_2$? To find out, let us write the initial and final states, $\ket{\psi(t_1)}$ and $\ket{\psi(t_2)}$, as two-particle states. Since the amplitude $\braket{\psi(t_2)}{\psi(t_1)}$ must come out equal to the sum of the amplitudes $A_1=\braket{C,D}{A,B}$ and $A_2=\pm\braket{D,C}{A,B}$, the appropriate two-particle states are
\begin{equation}
\ket{\psi(t_1)}=\frac1{\sqrt2}\Bigl(\ket{A,B}\pm\ket{B,A}\Bigr),\quad
\ket{\psi(t_2)}=\frac1{\sqrt2}\Bigl(\ket{C,D}\pm\ket{D,C}\Bigr).
\end{equation}
In other words, we must use (anti)symmetrized two-particle states. If we were to use $\ket{A,B}$ instead of the (anti)symmetrized product, we would introduce, in addition to the physically warranted distinction between ``the particle in $A$'' and ``the particle in $B$,'' the physically unwarranted distinction between the ``first'' or ``left'' particle and the ``second'' or ``right'' particle (in the expression $\ket{A,B}$). This would be justified if the particles carried ``identity tags'' corresponding to ``left'' and ``right,'' in which case we would be required to add probabilities, not amplitudes. If the distinction between ``the particle in $A$'' and ``the particle in $B$'' is the only physically warranted distinction, the distinction between the  ``left'' particle and the ``right'' particle must be eliminated, and this is achieved by (anti)symmetrization. The bottom line is that the absence of diachronic individuators entails the absence of synchronic individuators.

Quantum mechanics challenges us to think in ways that do not raise unanswerable questions. If we take for granted that space is an intrinsically differentiated expanse, we are led to ask the unanswerable question, ``Through which slit did the electron go?'' If we take for granted that at $t_1$ there are \emph{two} things, and that at $t_2$ there are again \emph{two} things, we are led to ask the unanswerable question, ``Which is identical with which?'' If instead we take the view that initially there is \emph{one} thing present in \emph{both} $A$ and~$B$, and that subsequently there is \emph{one} thing present in \emph{both} $C$ and~$D$, this unanswerable question can no longer be asked.

But then the following question arises. Because the two particles seen at $t_1$ or at $t_2$ lack individuating properties, they can only be \emph{numerically} distinct. But are they numerically distinct \emph{only} because they are observed in different regions of space? Ordinarily we tend to believe that there is another reason: two qualitatively identical objects situated in non-overlapping regions of space are numerically distinct because they are individual substances or, to put it more crudely, because they are made of stuff that can be parceled out. Yet the two particles, lacking as they do synchronic individuators, are numerically distinct for the \emph{sole} reason that they are observed in different regions of space. Considered without regard to their positions, they are therefore identical in the strong sense of numerical identity. What is present in the two regions is one and the same thing---a single entity that is ontologically anterior to spatial distinctions.

Moreover, the numerical identity of what is present in two regions need not be confined to so-called identical particles (particles of the same type). There is no compelling reason to believe that this identity ceases just because it ceases to have observable consequences when persistent distinguishing characteristics are present. What can be present in different places (i.e., possess different positions) can also possess different properties other than position. Nothing therefore stands in the way of the claim that, in and of itself, each fundamental particle---each particle that lacks internal structure---is numerically identical with every other fundamental particle.

There is an extensive literature on the subject of identity and individuality in quantum theory. See French~\cite{SEP-French} for an overview and French and Krause~\cite{FrenchKrause} for a comprehensive review. French sums up the situation by stating that quantum mechanics is ``compatible with two distinct metaphysical `packages,' one in which the particles are regarded as individuals and one in which they are not.'' In other words, there is an underdetermination between an ontology of quantum objects as individuals and an ontology of quantum objects as non-individuals. Esfeld~\cite{Esfeld2013}, however, denies that there is such an underdetermination because it is not ``a serious option to regard quantum objects as possessing a primitive thisness (haecceity) so that permuting these objects amounts to a real difference.''

What I shall explore in the remainder of this paper is the view (and its implications for the interpretation of quantum mechanics) that \emph{all} fundamental particles---not only the indistinguishable ones---are identically the same entity~$\cal E$.

\section{Particles: natural kinds}
\label{sec:particles}
Classical realism is indebted to two metaphysical assumptions. The first is that physical objects are \emph{substances}---independent carriers of primary qualities that can be defined in terms of monadic predicates. The second is that physical events and processes are completely determined by laws of nature in accordance with the principle of \emph{causality}~\cite[p.~27]{Falk}. Because the macroworld effectively conforms to the deterministic correlation laws of classical physics (except for the unpredictable transitions of value-indicating pointers), it lends itself to being described in causal terms. This makes it possible to sort macroscopic properties into bundles, and to attribute each bundle to a re-identifiable individual substance.

Two points should be noted here. The first is that the two assumptions are not independent. The possibility of sorting properties into bundles that are attributable to re-identifiable substances, only exists if the laws of nature are at least effectively deterministic. Because the correlations between macroscopic positions evince no statistical variations (value-indicating pointer transitions again excluded), macroscopic objects satisfy this condition. The second point is that the quantum-mechanical correlation laws themselves enable us to identify the domain in which causal stories about independent, re-identifiable objects can be told. What lies ``beneath'' this domain, in a sense that will shape up as we proceed, does not conform to this narrative mode:
\begin{quote}
quantum mechanics is neither compatible with the traditional concept of substance (that is, the principle of attributing properties to property carriers) nor with the principle of causality in its usual application to individual systems and processes~\cite[p.~28]{Falk}.
\end{quote}
For Falkenburg this is the reason---``which is hardly recognized in recent debates on quantum theory''~\cite[p.~29]{Falk}---why making physical sense of the mathematical formalism of quantum mechanics is so hard. 

The question then is: what \emph{causes} a detector to click? What is clear right away is that the quantum-mechanical correlation laws cannot provide the answer. The outcome represented by the span ${\cal A}\cup{\cal B}$ of two orthogonal subspaces of a Hilbert space (or by the corresponding projector) can be certain even if neither of the respective outcomes represented by ${\cal A}$ and ${\cal B}$ is certain. If the probabilities assigned to these outcomes are less than~1 while the probability assigned to ${\cal A}\cup{\cal B}$ is~1, it is certain that either of the two outcomes will be found even though neither outcome is certain to be found. Whence this certainty? 

The answer lies in the fact that quantum-mechanical probability assignments are invariably made on the more or less tacit assumption that a measurement is successfully made: there is an outcome. Quantum-mechanical probabilities are conditional not only because they are assigned on the basis of actual measurement outcomes but also because they are assigned on the assumption that an outcome will be obtained. While quantum mechanics presupposes value-indicating events, its doubly conditional probability assignments do not allow us to formulate causally sufficient conditions for the occurrence of such events~\cite{UlfBohr,Mohrhoff-clicks}. The efficiency of a detector can of course be measured, but it cannot be calculated from first principles without making use of parameters (particularly coupling constants) which themselves have to be experimentally determined.%
		\footnote{Question: If a photon passes a beam splitter, is it not certain to be detected in either of the beams? Answer: What is certain is that the photon will be detected if each beam enters a \emph{perfect} (100\% percent efficient) detector. Perfect detectors, however, do not exist. A statement involving perfect detectors is equivalent to a conditional statement: the photon will be detected in either beam \emph{if} a measurement designed to determine the beam taken by the photon is successfully made.}

This brings us to the question: what exactly do we speak of when we speak of a ``particle''? For one thing, we do not speak of something that makes detectors click. If a detector clicks, it is not because a particle has entered its sensitive region. Rather, the click is the reason---\emph{one} reason---why the property of being inside the detector's sensitive region can be attributed to something which, for want of a better word, we refer to as a ``particle.''

Another reason is the existence of conservation laws. If we perform a series of position measurements, and if each measurement yields exactly one outcome (i.e., if each time exactly one detector clicks), we have evidence of a conservation law. If each time exactly two detectors click, we have evidence of the same conservation law, the conserved quantity being the (maximum) number of possible simultaneous detector clicks. If this is the only conservation law in force, then we cannot interpret the number of simultaneous clicks as the number of re-identifiable individuals. What we can infer is the existence of a single ``something'' to which a fixed number of positions can be simultaneously attributed---if and when its positions are measured. If in addition there are separate conservation laws for different types of clicks (say, baryon clicks and lepton clicks), then we \emph{could} infer the existence of two things each with a fixed number of measurable positions, though it would clearly make more sense to infer the existence of a single object to which a fixed number $N_B$ of baryon positions and a fixed number $N_L$ of lepton positions can be simultaneously attributed (if and when its positions are measured).%
		\footnote{This argument is for illustrative purposes only. $N_B$~and $N_L$ are not the (approximately) conserved baryon and lepton numbers.}

This raises the question: what makes a click of a particular type $T$ (say, an electron click) a click of type~$T$? A single click does not announce what type of particle has been detected. The type to which a detection event belongs has to be inferred from a sequence of detection events---an optically recorded sequence of events forming a track in a bubble, cloud, spark, or streamer chamber, or a sequence of events electronically acquired by using a wire or drift chamber or a silicon detection system.  We classify detection events according to the type of particle detected, and we classify particles on the basis of sequences of detection events. A sequence of detection events makes it possible to measure such quantities as the radius of curvature of a particle's track in a magnetic field, a particle's time of flight, a particle's kinetic energy, and a particle's energy loss through ionization and excitation. Three of these four measurements are in principle sufficient to positively identify the particle type~\cite[Chapter~9]{GS}, which in turn makes it possible to classify the individual detection events.%
		\footnote{While neutral particles cannot be inferred directly from particle tracks, they can be inferred indirectly from their interactions with charged particles, on the basis of conservation laws.}

According to Falkenburg~\cite{Falk}, subatomic particles are ``collections of empirical properties which constantly go together or bundles of properties which repeatedly appear together'' (p.~221). Their properties `` are tied together to property bundles with the status of natural kinds'' (p.~259). Although these property bundles ``are only individuated by the experimental apparatus in which they are measured or the concrete quantum phenomenon to which they belong'' (p.~206), ``there \emph{is} an entity which appears as a stable bundle of properties in the phenomena'' (p.~259, original emphasis).%
	\footnote{Quarks ``appear in the phenomena'' in a less direct manner than the other standard-model fermions. They appear ``as dynamically discontinuous constituent parts of localizable bound systems''~\cite[p.~262]{Falk}.}

There is indeed something which appears in the phenomena. The question is only: is what appears (a)~a stable bundle of properties, or is it (b)~something that appears \emph{as} a stable bundle of properties? If one set of particles (for example, a probe and a target particle in a high-energy scattering experiment) can be transformed into a different set of particles, the answer must be~(b). What appears is not a natural kind but something that appears now as one set of particles (i.e., tokens of natural kinds) and subsequently as another such set. What appears is that single entity~$\cal E$ introduced in the previous section, which can be present now under the form of two tokens of the natural kinds $T_A$ and~$T_B$, respectively, located in regions $A$ and~$B$, and subsequently under the form of two tokens of the natural kinds $T_C$ and~$T_D$, respectively, located in regions $C$ and~$D$.

\section{Particles: localizability}
\label{sec:lp}
Hegerfeldt~\cite{Hf74,Hf98,Hf2001} has shown that in a relativistic quantum theory superluminal spreading of wave functions must occur if (i)~there are states with localized particles and (ii)~there is a lower bound on the system's energy. (A particle is said to be localized at $t_0$ if it is prepared in such a way as to ensure that it will be found with probability~1, by a measurement performed at~$t_0$, within a finite region of space.) The conclusion is that it is impossible to localize a particle even for an instant.
 
Malament~\cite{Malament} has proved a theorem which establishes that in relativistic quantum mechanics particles cannot be completely localized in spatial regions with sharp boundaries. Clifton and Halvorson~\cite{CH} have strengthened Malament's result by deriving a theorem which shows that there is no relativistic quantum mechanics of unsharply localized particles either. Quantum theory engenders a fundamental conflict between relativistic causality and localizability, so that particle talk is ``strictly fictional'': 
\begin{quote}
The argument for localizable particles appears to be very simple: Our experience shows us that objects (particles) occupy finite regions of space. But the reply to this argument is just as simple: These experiences are illusory! Although no object is strictly localized in a bounded region of space, an object can be well-enough localized to give the appearance to us (finite observers) that it is strictly localized.
\end{quote}
What Hegerfeldt, Malament, and Clifton and Halvorson have established is analogous in some respects to the non-objectification theorems proved by Mittelstaedt \cite[Sect.~4.3(b)]{Mittelstaedt} and the insolubility theorem for the objectification problem due to Busch et al.~\cite[Sect.~III.6.2]{BLM}. While non-relativistic quantum mechanics cannot account for the measurement outcomes which it presupposes and serves to correlate, relativistic quantum field theory cannot account for the particle detections which it presupposes and serves to correlate. The latter theory provides us with conditional statements of the following form: if a set of particles ${\cal P}_i$ with momenta $p_i$ come together in a scattering event, then such is the probability with which a set of particles ${\cal Q}_k$ with momenta $q_k$ will emerge from the event. The theory requires that in-states be prepared and out-states be observed, but it leaves the operational implementation to the experimenters. Experimenters use a generalized version of Bohr's correspondence principle to identify these states with the particle types and momenta they obtain by analyzing particle tracks, and it is these data that the theory serves to correlate.%
		\footnote{According to Falkenburg~\cite[p. XII]{Falk}, ``quantum mechanics and quantum field theory only refer to individual systems due to the ways in which the quantum models of matter and subatomic interactions are linked by semi-classical models to the classical models of subatomic structure and scattering processes. All these links are based on tacit use of a generalized correspondence principle in Bohr's sense (plus other unifying principles of physics).'' This generalized correspondence principle, due to Heisenberg~\cite{Heisenberg}, serves as ``a semantic principle of continuity which guarantees that the predicates for physical properties such as `position', `momentum', `mass', `energy', etc., can also be defined in the domain of quantum mechanics, and that one may interpret them operationally in accordance with classical measurement methods. It provides a great many inter-theoretical relations, by means of which the formal concepts and models of quantum mechanics can be filled with physical meaning''~\cite[p.~191]{Falk}.}

Whether or not particles are localizable depends on our answer to the question: localizable relative to what? The most basic axiom of field theory---so basic that it is rarely explicitly stated---postulates the existence of a spatiotemporal manifold~$\cal M$. What Hegerfeldt, Malament, and Clifton and Halvorson have shown is that a particle cannot be in a state in which the probability of finding it within any finite spatial region of $\cal M$ equals~1. $\cal M$, however, is not where experiments are performed. The possible outcomes of position measurements are not defined relative to a completely differentiated spatial background. Attributable positions are defined by the sensitive regions of detectors, which also cannot be localized in any finite spatial region of $\cal M$. What is strictly fictional, therefore, is the spacetime manifold postulated by quantum field theory. What Clifton and Halvorson have shown is not that there are no localizable particles but that this manifold is not localizable relative to the positions that particles can possess.

\section{The shapes of things}
\label{sec:Shapes}
According to Ladyman and Ross~\cite[p.~4]{LadyRoss}, ``it is no longer helpful to conceive of either the world, or particular systems of the world that we study in partial isolation, as `made of' anything at all.'' I agree. If things are made of anything at all, they are made of a single entity~$\cal E$, which is both ontologically anterior to and capable of entertaining reflexive spatial relations. The interesting question is not what \emph{things} are made of but what \emph{forms} are made of. Forms are made of (reflexive) spatial relations, or so I shall now argue.

The idea most commonly associated with the concept of a spatial form is that of a boundary. It is closely related to the concept of matter in ancient philosophy, which gave rise to the conundrum about the divisibility of matter by \emph{cutting}, i.e., by introducing delimiting surfaces. (The literal meaning of the Greek word \emph{atomos} is ``uncuttable.'') This idea, moreover, appears to have neurobiological underpinnings. The manner in which the brain appears to process visual information guarantees that the result---the visual world---is a world of objects whose shapes are bounding surfaces~\cite{Hubel95,Enns04,Hoffman98}. The tension of contrast between shapes qua (definite) bounding surfaces and shapes qua sets of (more or less indefinite) spatial relations is perhaps the most overlooked significant difference between classical and quantum conceptions of reality.

Classically conceived, the shapes of things resolve themselves into the shapes of their constituents and the spatial relations between them. Since classical relativity rules out the existence of rigid bodies, the constituents have to be either elastic or pointlike. If elasticity is understood in terms of variable distances between constituents, there must be ultimate constituents, and they must be pointlike.

Fundamental particles are often said to be pointlike, but not in the same classical sense, or else for the wrong reason. A particle is said to be pointlike if it lacks internal structure.%
		\footnote{It may be asked whether there are particles that are fundamental in this sense. In 1998, the Elementary-Particle Physics Panel of the U.S. National Research Council~\cite[p.~23]{EPPP} stated that ``[t]he question is still open experimentally, but theory and experiment are pointing more than ever before to the possibility that we have discovered the `ultimate constituents'.'' As recently as 2013, Nicolai~\cite{Nicolai} affirms that ``there is not a shred of a hint so far that would point to an extended structure of the fundamental constituents of matter (quarks, leptons and gauge bosons).''}
A particle's lack of internal structure can be inferred from the scale-invariance of its effective cross-section(s) in scattering experiments with probe particles that are themselves pointlike (in this sense). Since spatial structures are not measurable below the de Broglie wavelength of the probe particles, no scattering experiment can provide evidence of absence of internal structure, let alone evidence of a literally pointlike form. 

A more insidious reason for conceiving of particles as pointlike is rooted in the fact that relativistic conservation laws must be local. To ensure that for every conserved quantity the total associated with the incoming particles in a scattering experiment equals the total associated with the outgoing particles, the conserved quantities must ``flow locally'' from the incoming particles to the outgoing particles. If scattering amplitudes are calculated using Feynman diagrams in position space, the locality of this flow is guaranteed by the unbroken lines representing propagators and the pointlike aspect of the vertices at which they meet. A literal physical interpretation of these computational aids may then lead to the conclusion that particles interact with each other only when they are in the same place, i.e., when their positions relative to each other are sharp.
 
Transition amplitudes, however, tell us nothing about what actually takes place during a transition, be it an atomic transition or the transition from a given set of in-states to a given set of out-states.%
		\footnote{``The S-matrix\dots gives transition probabilities which correspond to measurable relative frequencies. But it treats the scattering itself as a black box\dots. Feynman diagrams\dots have no literal meaning. They are mere iconic representations of the perturbation expansion of a quantum field theory. They make the calculations easier, but they do not represent individual physical processes.''~\cite[p.~131--132]{Falk}}
We may conceive of a perturbation expansion as a sum over all possible ways in which a transition can take place, but the very fact that we sum amplitudes associated with distinct graphs implies (via YAIP) that the distinctions we make between the graphs cannot be considered objective. (The same applies to the propagators, which are sums over spacetime paths,%
		\footnote{In field theories, the sums over spacetime paths are implicit in the representation of particles by fields, inasmuch as the fields are solutions of the dynamical equations one obtains by summing over spacetime paths.}
and to the vertices, which are integrated over spacetime, and thus as non-local as it gets.) Because the distinctions we make between the alternatives represented by the graphs cannot be considered objective, the alternatives act as one, giving rise to collective effects like the Lamb shift, the masses of nucleons, or the momentum-dependence of masses and coupling parameters.

In sum, the idea that a fundamental particle is a literally pointlike object is unwarranted on both theoretical and experimental grounds. In addition, it is inconsistent with the incomplete spatial differentiation of the physical world, inasmuch as it would imply that something is present at some $x$ while nothing is present in the infinitesimal neighborhood of $x$, which is possible only in a world that is spatially differentiated ``all the way down.'' Nothing therefore stands in the way of the claim that a fundamental particle is a literally formless object; on the contrary, everything speaks for it.

One conclusion reached in Sect.~\ref{sec:tse} was that \emph{if} we think of space as a substantial expanse, we must think of it as undifferentiated. If fundamental particles are formless, a different conception of physical space suggests itself. The notion of a literally pointlike object entails the existence of a spatial expanse in which it is situated. If there are no pointlike objects, there is no need to conceive of such an expanse. Space may be conceived as containing---in the set-theoretic sense of containment---nothing but spatial relations, the relata being formless objects and ``composites'' thereof. Being spatially extended would then be not a property of a substantial expanse but a quality shared by all spatial relations. If we adopt this conception of space, the truism that the universe lacks a position because it lacks \emph{external} spatial relations acquires a fitting complement: a fundamental particle lacks a form because it lacks \emph{internal} spatial relations.

This patently relationist conception of space goes farther in relationism---the doctrine that space and time are a family of spatial and temporal relations holding among the material constituents of the universe---in that it also affirms that the ``ultimate'' material constituents are formless. If space consists of relations between formless relata, the latter are not contained in space. What makes it possible for them to be detected at different locations is the spatial relations that hold among them. This makes it possible to go farther still and assert that, in and of itself, each fundamental particle is numerically identical with $\cal E$, and thus with every other fundamental particle, so that the relations that hold between fundamental particles are reflexive.%
		\footnote{Those who wish to conceive of space (or spacetime) as a self-existent expanse may do so---on condition that they conceive of it as undifferentiated. What is differentiated is its material ``content.'' Because this is not differentiated all the way down, the multiplicity inherent in the set-theoretic conception of space (or spacetime) cannot be considered objective. In other words, while 
substantivalism with regard to an undifferentiated spatiotemporal expanse is defensible, substantivalism with regard to a set-theoretically conceived ``continuum'' is not.}

The form defined by a set of spatial relations is an abstract concept. The form of a bipartite object---for instance that of a hydrogen atom if the proton's structure is ignored---consists of a single relative position~$\mathbf r$. It can be described by a probability distribution over the possible outcomes of a measurement of~$\mathbf r$. If we have measured, say, the atom's energy, its total angular momentum, and one component of its angular momentum, then we know how to calculate this probability distribution. (Since this description does not assume that the measurement is actually made, we need not be overly worried by the nonexistence of suitable detectors.)

The form of an object with $N>2$ components consists of $N(N{-}1)/2$ spatial relations. While the positions of $N{-}1$ components relative to the remaining component can be described with the help of a single probability distribution over a $3(N{-}1)$-dimensional abstract space, the relative positions between these $N{-}1$ components can only be described in terms of correlations between the outcomes of measurements of their positions.

The abstractly defined forms of nucleons (bound states of quarks), nuclei (bound states of nucleons), atoms (bound states of nuclei and electrons), and molecules (bound states of atoms) ``exist'' in probability spaces of increasingly higher dimensions. At the molecular level of complexity, however, an entirely different kind of form comes into being: a 3-dimensional form that can be visualized, not as a distribution over a 3-dimensional probability space, like the abstract form of a hydrogen atom in a stationary state, but \emph{as it is}. I am of course speaking of the atomic configurations of molecules. What contributes to making these configurations visualizable is that the indefiniteness of the distance $d$ between any pair of bonded atoms, as measured by the standard deviation of the corresponding probability distribution, is significantly smaller in general than the mean value of~$d$.

If classical properties or behaviors emerge, it is not from some mystical domain of potentialities, nor by a dynamical process, but in the conceptual transition from the compositionally simple to the complex. If there is a quantum-classical boundary, it is molecules that straddle it. There is something on the classical side, to wit, their atomic configurations, which change slowly, while the electron wave functions follow adiabatically. Only molecules consisting of very few atoms are known to occur in energy and angular momentum eigenstates~\cite[p.~99]{Joosetal2003}.

It is now widely recognized that environmental decoherence~\cite{Zurek2003,Joosetal2003,Schlosshauer07} contributes significantly to the emergence of classicality at the molecular level. Decoherence induced by the scattering of particles---dust grains, air molecules, thermal and microwave photons, even solar neutrinos---is seen as responsible for, \emph{inter alia}, the handedness of sugar molecules, the parity of ammonia molecules, the definite orientations of larger molecules, and the tertiary structures of DNA-molecules and proteins. It should be borne in mind, though, that decoherence only results in a quantum system's getting entangled with the environment; the resulting mixed state does not admit of an ignorance interpretation---\emph{unless} ``environment'' stands for the macroworld. While the indefiniteness of a system observable can infect its microscopic environment, giving rise to (synchronic) correlations between the possible outcomes of measurements performed on the system and this environment, it cannot infect the macroworld, for if it did, there would be no measurement outcomes, we could not speak of correlations between measurement outcomes, and the term ``entanglement'' would be undefined. If, on the other hand, the environment includes the macroworld, or if ``environment'' stands for it, what happens instead is that the definiteness of the macroworld reproduces itself in the genuinely quantum domain.

This spreading of macroscopic definiteness into the genuinely quantum domain is what makes it possible to extrapolate the classical understanding of objects in terms of interacting parts to about the scale of large molecular structures, as any textbook of molecular biology amply demonstrates. Below this scale, descriptions in terms of composition and parts go astray, inasmuch as there the ``parts'' lose their separate identities, and the state of a ``composite'' object generally ceases to be determined by the states of its ``parts.'' (Whenever a pure state can be assigned to a composite object, state separability fails, either because pure states cannot be assigned to the parts, or because the density operators of the parts do not uniquely determine the density operator of the whole.)

\section{An apparent logical circle}
\label{sec:lc}
If kinematical properties are attributable to microphysical objects only if (and when) their possession is indicated by macroscopic objects, then the popular notion that macroscopic objects are made up of microphysical ones leads to a vicious circle. To resolve it, we need not abandon the view that microphysical objects contribute to a macroscopic object's being what it is. We only need to understand how microphysical objects  contribute to---what role they play in---making a macroscopic object what it is.

One of the salient differences between classical and quantum accounts is that the latter interpose an extra level of description between a physical system and its properties. A classical system can be described in terms of properties that exist whether or not their presence is indicated by measurements. Quantum systems, by contrast, are described (a)~in terms of correlations between the possible outcomes of measurements (irrespective of whether they are made) and (b)~by the outcomes of measurements that \emph{are} made. 

When a particle or atom is described in isolation, it is described in terms of (diachronic) correlations between outcomes of measurements to which the particle or atom can be subjected. An electron in isolation is described by an irreducible representation of a symmetry group, which defines a propagator or a dynamical equation, which defines correlations between detection events. Strictly speaking, this is a description not of the natural kind \emph{electron} but of the natural kind \emph{electron in isolation}. While this description involves no actual measurements and does not individualize, it makes it possible to identify electrons by identifying particle tracks as electron tracks. It is the electron tracks that individualize, and that make it possible to attribute to the individual electron both a sequence of more or less definite positions and a sequence of more or less definite momenta (by connecting successive detection events).

When electrons are among the so-called constituents of an atom or a molecule, on the other hand, we are no longer dealing with the natural kind \emph{electron in isolation}. Such electrons are not individualized, nor can kinematical properties be attributed to them (since none are measured). It is the isolated atom or molecule that then is described (a)~in terms of (diachronic) correlations between the possible outcomes of measurements to which it can be subjected (irrespective of whether they are made) and (b)~by the outcomes of measurements that \emph{are} made. But if electrons need no kinematical properties to contribute to making an isolated atom or molecule what it is, they need no kinematical properties to contribute to making a macroscopic object what it is. Hence, no vicious circle ensues.

A circularity exists, but it is revealing rather than vicious. Landau and Lifshitz~\cite[p.~3]{LL77} point to it when they observe that ``quantum mechanics\dots contains classical mechanics as a limiting case, yet at the same time it requires this limiting case for its own formulation.'' So does Redhead~\cite{Redhead1990} in stating this implication of the Copenhagen interpretation: ``In a sense the reduction instead of descending linearly towards the elementary particles, moves in a circle, linking the reductive basis back to the higher levels.''

In resisting attempts at explaining things in compositional terms, i.e., by explaining the unity of a whole in terms of a multiplicity of parts, quantum mechanics calls for an explanatory concept that proceeds in the opposite direction: from the unity of an intrinsically undifferentiated entity~$\cal E$ to the multiplicity of the macroworld, by means of an atemporal process of differentiation. If there is a distinctly quantum domain, it is not a domain of constituent objects with intrinsic properties but a domain in which---or, rather, an aspatial and atemporal dimension across which---this transition from unity to multiplicity takes place. While atoms and subatomic particles are instrumental in this transition, their instrumentality cannot be understood in compositional terms. Ultimately there is but one constituent, to wit,~$\cal E$.

One can arrive at this conclusion by observing that the properties of the macroworld allow themselves to be bundled into separate objects only up to a point, and that physical space likewise allows itself to be partitioned into disjoint regions only up to a point. If we go on dividing a material object, we reach a point where the distinctions we make between component parts cease to correspond to anything real, a point where the ``component parts'' become identical in the strong sense of numerical identity (Sect.~\ref{sec:Identity}). And if we keep partitioning physical space, we reach a point where the distinctions we make between regions of space cease to correspond to anything real, a point where physical space becomes an intrinsically undifferentiated expanse (Sect.~\ref{sec:tse}). We are free to think not only of that which every fundamental particles intrinsically is but also of this intrinsically undifferentiated expanse as aspects of~$\cal E$. By entering into reflexive spatial relations, $\cal E$~gives rise to (i)~what looks like a multiplicity of relata whenever the reflexive quality of the relations is not taken into account, and to (ii)~what looks like a substantial expanse if the spatial quality of the relations is reified.   

On the way from the unity of $\cal E$ to the macroworld we encounter increasingly differentiated structures such as identical particles, non-visualizable atoms, and partly visualizable molecules. We encounter systems that are described in terms of correlations across probability spaces of increasingly higher dimensions. At the molecular level of differentiation, macroworld-induced decoherence gains traction, and (the physical aspect of) what Sellars~\cite{Sellars} has called ``the manifest image of the world'' comes into view. 

The previously mentioned (revealing rather than vicious) circularity ensues because the transition from unity to multiplicity goes through stages of increasing definiteness and distinguishability. To describe what is indefinite and indistinguishable, we resort to probability distributions over events that are definite and distinguishable, and such events only exists in the macroworld. What is instrumental in the transition can only be described in terms of the final result, the macroworld.

\section{Manifestation}
\label{sec:Manifestation}
We need a name for this new explanatory concept---this transition from unity to multiplicity across an aspatial and atemporal dimension, this self-differentiation of $\cal E$ through the adoption of reflexive spatial relations---and Sellars's distinction between the manifest and scientific images of the world suggest what it should be.

For Sellars~\cite{Sellars}, the manifest image is, broadly speaking, the framework in terms of which we ordinarily explain our world (including ourselves). It seeks to establish the correlations that we observe, without trying to explain them in terms of the theoretical postulates of scientific theories. While the manifest image thus is devoid of the theoretical posits that populate the scientific image, the scientific image, according to Sellars, lacks the ``categories pertaining to man as a person,'' such as intentions, thoughts, and appearances.

To make a distinction between the two images, however, it is not necessary to bring up issues pertaining to the philosophy of mind. The distinction can also be made in the inanimate world,  as illustrated by Eddington's two tables~\cite[pp. ix--x]{Eddington}. The relevance of Sellars's distinction to the interpretation of quantum mechanics was noted by Maudlin~\cite{Maudlin}:
\begin{quote}
Insofar as one uses quantum theory to construct a Scientific Image of the world, that Image will include a wave function. Further, if everything that exists must be represented in the Scientific Image, then there should be a wave-function of everything, i.e., of the whole universe. Now suppose that the wave function is all there is in the Scientific Image (i.e., suppose the wave function is complete). And suppose further that the wave function never collapses. Then it is very difficult to see how to make contact between the Scientific and Manifest Images of the world. As is famously illustrated by the example of Schr\"odinger's cat, the wave function's behavior is not even loosely isomorphic to the Manifest behavior of the cat. The Scientific Image must be amended or expanded if we are to find some \emph{doppelg\"anger} of the Manifest Image in it.
\end{quote}
According to Maudlin, the fundamental interpretive problem is exactly the problem of connecting the scientific image with the manifest. It seems evident that the points of contact must be positions---the definite positions of things in the manifest image and the positions of certain quantum objects in the scientific. If we wanted to identify the momenta of manifest things with the momenta of certain quantum objects, we would not know where to begin. As Maudlin observes, there is ``simply no obvious way to sketch any isomorphism between a world of particles which only have momentum and the world as we experience it.''

I have argued for a way to amend a misconceived scientific image, a way that rejects two extreme solutions of the fundamental interpretive problem---a radical empiricism or metaphysically sterile instrumentalism on one hand, and on the other the reification of the wave function and/or of the spacetime points on which it functionally depends. The empiricist is right in that quantum states are probability algorithms, the  instrumentalist is right in that the quantum-mechanical probability calculus serves to correlate value-indicating events, but both are wrong when they assert that correlations between observations or macro-events are all there is. There is this single, intrinsically undifferentiated entity $\cal E$, which enters into indefinite and reflexive spatial relations. There are the quantum-mechanical correlation laws, which together with these relations give rise to structures such as atoms, which are instrumental in \emph{manifesting} (i.e., making manifest) a world of macroscopic objects, whose positions are real \emph{per se}. Realists would be right in positing these things but they are wrong in transmogrifying calculational tools into physical states evolving in an intrinsically and completely differentiated spacetime.

By claiming that the macroworld results from a self-differentiation of~$\cal E$, or that~$\cal E$ manifests the macroworld by entering into reflexive relations, I posit a so far unrecognized kind of causality---unrecognized, I believe, within the scientific literature albeit well-known to metaphysics, for the general philosophical pattern of a single world-essence manifesting itself as a multiplicity of physical individuals is found throughout the world. Some of its representatives in the Western hemisphere are the Neoplatonists, John Scottus Eriugena, and the German idealists. The quintessential Eastern example is the original (pre-illusionist) Vedanta of the Upanishads~\cite{Phillips,Aurobindo}. This understands the world as the manifestation of an ultimate reality, which is related to its manifestation in three mutually irreducible ways: as its constituent substance, as its containing consciousness, and as a pure quality/delight as yet undifferentiated into its subjective and objective aspects.%
		\footnote{Sri Aurobindo~\cite{Aurobindo,Mohrhoff-CQM} offers a detailed account of how that ultimate reality comes to take on these essential aspects.}

To my mind, the Vedantic world conception in particular offers a promising framework for a full resolution of the tension between the scientific and manifest images, in the broader sense intended by Sellars. After all, if the world is manifested \emph{by} something or someone, it is also manifested \emph{to} something or someone. The Vedantic concept of manifestation both implies a duality of mind and matter and transcends it, inasmuch as that \emph{by} which the world is manifested is ultimately one with that \emph{to} which the world is manifested.%
		\footnote{``Hence I am God Almighty,'' Schr\"odinger concludes in the Epilogue to his classic, \emph{What is Life}?~\cite{SchrLife}, in which he makes explicit reference to Vedanta and the Upanishads.}
Needless to say, a discussion of this identity and of what it entails far exceeds the scope of the present paper~\cite{Mohrhoff-CQM}.%
		\footnote{I venture, however, to say this much: 	
Ever since Leibniz, philosophers have argued that all physical properties are relational or extrinsic, and none are in a fundamental sense non-relational or intrinsic. As was pointed out by Russell~\cite{Russell}, this offers the possibility of locating the ``categories pertaining to man as a person''~\cite{Sellars} in the intrinsic properties of the relata that bear the relational physical properties. It also makes it possible to think of ultimate reality's essential aspect of quality/delight as the intrinsic nature of~$\cal E$. A case can therefore be made for a reversal of metaphysical reductionism: what matters metaphysically is first and foremost what is manifested, or what the manifest image contains; particles are but means to manifest it (and so, presumably, are brains).}

If the spatiotemporal differentiation of the world is incomplete, and if, as a consequence, the interpretation of quantum states as evolving physical states is ruled out, then the causality of the atemporal process of manifestation is the only kind of causality that is applicable to the distinctly quantum domain. Because the quantum-mechanical correlation laws are (effectively) deterministic only within the macroworld, the temporal concept of causation, which links states or events across time or spacetime, has meaningful application only in the macroworld.%
		\footnote{Ladyman and Ross~\cite[pp.~258, 280]{LadyRoss} concur: ``the idea of causation has similar status to those of cohesion, forces, and things. It is a concept that structures the notional worlds of observers\dots. Appreciating the role of causation in this notional world is crucial to understanding the nature of the special sciences, and the general ways in which they differ from fundamental physics\dots. There is no justification for the neo-scholastic projection of causation all the way down to fundamental physics and metaphysics.''}
It plays no role in its manifestation. It is part of the world drama, but it does not take part in setting the stage for it.%
		\footnote{One should also bear in mind that quantum theory's doubly conditional probability assignments do not allow us to formulate causally sufficient conditions for value-indicating events (Sect.~\ref{sec:particles}). While the indicated values of observables (as well as the times at which they are possessed) can be considered objective, causal relations between them cannot.}

The causality of the atemporal process of manifestation also has teleological aspects. It manifests a world---the macroworld---whose properties allow themselves to be sorted into (causally) evolving bundles of re-identifiable individual substances. This opens the door to ``anthropic'' arguments~\cite{BarrowTipler,Rees,Bostrom}. What does it take to manifest such a world? At a minimum, such a world must contain objects that have spatial extent and are (relatively) stable~\cite{Mohrhoff-QMexplained}. To manifest such objects by means of finite numbers of spatial relations (both relative positions and relative orientations), these relations (as well as the corresponding relative momenta) have to be indefinite, uncertainty relations must hold, the relata have to be fermions, and there are further conditions that must be satisfied~\cite[Chapters 8 and 22]{Mohrhoff-book}. 

The question then is, how do we deal with indefinite observables in a rigorous manner? We do so by assigning probabilities to the possible outcomes of measurements of such observables. This is why the irreducible empirical core of quantum mechanics is a probability calculus. And how are the possible outcomes defined? They are defined by macroscopic instruments (Sect.~\ref{sec:tse}). By realizing properties, a measurement apparatus makes properties available for attribution, and by indicating a particular outcome, it warrants the attribution of a particular property. This is why the events to which quantum mechanics assigns probabilities are measurement outcomes. In short, the manifestation of a world in which causal stories about independent, re-identifiable objects can be told, entails the validity of something very much like quantum mechanics. 

What more does the manifestation of a world which conforms to the classical narrative mode entail concerning the laws and the natural kinds or structures that are instrumental in its manifestation? If one also requires that quantum theory be discoverable and/or testable (without necessarily requiring that anyone be around to discover and/or test it), then a case can be made that all of the four empirically known ``forces'' are needed~\cite{Mohrhoff-QMexplained,Mohrhoff-book,Mohrhoff-justso}. The strong force (QCD) results in particles of different types: three types of colored quark and eight types of gluon. These give rise to a first layer of bound states (hadrons), among them the nucleons, which give rise to a second layer of bound states (atomic nuclei). Adding the weak force (necessary for stellar nucleosynthesis and the release of nuclei heavier than helium into the interstellar medium) amounts to multiplying the quark types by two (flavors) and to introducing another fermion doublet (the leptons) as well as three vector bosons.%
	\footnote{Why there are $3 \times 2$ flavors remains something of a mystery.}
Adding the electric force (QED) by introducing the photon and attributing electric charges to quarks, electrons, and two vector bosons gives rise to another two layers of bound states (atoms and molecules), and adding gravity produces a final layer of (gravitationally) bound states. If the ``stagflation'' of theoretical particle physics since the mid-1970's is an indication, it is difficult in the extreme to come up with something that beats the economy of the standard model%
		\footnote{According to Wilczek~\cite[p.~164]{Wilczek}, ``Standard model is a grotesquely modest name for one of humankind's greatest achievements.''}
plus general relativity in accounting for the existence of a world that conforms to the classical narrative mode.

It may be instructive to compare the manifestation of the macroworld with the classical philosophical concept of the emergence of the Many out of a One. In classical philosophy, this emergence was conceived as running parallel to predication: an immaterial essence or predicable universal becomes instantiated as an impredicable material individual. This instantiation, moreover, was conceived in the framework of a Platonic-Aristotelian dualism, which postulates an instantiating medium (matter and/or space) in or by which the essences or universals get instantiated.%
		\footnote{Ladyman and Ross~\cite[p.~155]{LadyRoss} ask their readers ``to consider whether the main metaphysical idea we propose, of existent structures that are not composed out of more basic entities, is any more obscure or bizarre than the instantiation relation in the theory of universals.'' It is not.}
 The manifestation of the macroworld, by contrast, requires no separate medium and implies no dualism. All that is required is the realization of spatial relations. $\cal E$~may be said to manifest the macroworld within itself---after all, the macroworld is manifested with the help of \emph{reflexive} relations---rather than in something other than itself. 

In her highly commendable book \emph{Particle Metaphysics}~\cite{Falk}, a wholesome antidote to mathematical literalism from which I have quoted several times, Brigitte Falkenburg defends a relational account of subatomic reality resulting from a top-down approach:
\begin{quote}
The opposite bottom-up explanation of the classical macroscopic world in terms of electrons, light quanta, quarks, and some other particles remains an empty promise. Any attempt at constructing a particle or field ontology gives rise to a non-relational account of a subatomic reality made up of independent substances and causal agents. But any known approach of this type is either at odds with the principles of relativistic quantum theory or with the assumption that quantum measurements give rise to actual events in a classical world. As long as the quantum measurement problem is unresolved, an independent quantum reality is simply not available. (p.~339)
\end{quote}
How does Falkenburg's relational top-down account differ from that proposed by the present author? We both reject bottom-up explanations, and we both have, consequently, characterized our respective approaches as being ``top-down,'' albeit in different senses. For Falkenburg, the macroworld is on top, and the downward direction indicates the epistemological dependence on it of subatomic reality: 
\begin{quote}
to our present knowledge subatomic reality is not a micro-world on its own but a part of empirical reality that exists relative to the macroscopic world, in given experimental arrangements and well-defined physical contexts outside the laboratory. (p. 340)
\end{quote}
The view put forward in the present paper puts on top a single, self-differentiating yet intrinsically undifferentiated  entity~$\cal E$, and the downward direction indicates the atemporal process of self-differentiation by means of indefinite and reflexive spatial relations. This view does not conflict with Falkenburg's account but rather complements it. As I see it, the relation between the macroworld and subatomic reality is not merely a one-way epistemological relation. It is this, inasmuch as we appear to have no choice but to describe what is instrumental in the process of manifestation in terms of its final outcome, the macroworld. But in addition to this epistemological dependence I infer (or postulate) an ontological dependence of the macroworld on~$\cal E$, in which subatomic reality plays an instrumental role.

In the passage just quoted Falkenburg states that no independent quantum reality is available as long as the quantum measurement problem is unresolved. Nor has such a reality been made available by the resolution of the problem in Sects.~\ref{sec:mmp} through \ref{sec:tse}. Subatomic reality's epistemological dependence on the macroworld remains, and although an independent reality has become available, it is not a reality made up of independent substances and causal agents. On the contrary, applying the concepts of substance and causality to a micro-world on its own would put the cart in front of the horse, for it is the quantum-mechanical correlation laws that tell us where these concepts have meaningful application. Instead of playing an instrumental role in the manifestation of the macroworld, these concepts have meaningful application only within the macroworld.

\section{Relations all the way down?}
\label{sec:OSR}
Today structural realism is considered by many to be the most defensible form of scientific realism~\cite{SEP-Ladyman}. The \emph{epistemic} structural realist holds that all we can know is structure, but it is the structure of an unknowable realm of individuals and their intrinsic natures. In the view of Ladyman and Ross~\cite{LadyRoss}, the naturalistic stance entails that talk of unknowable individuals and intrinsic natures is idle metaphysics. \emph{Ontic} structural realism (OSR), introduced by French and Ladyman~\cite{FrenchLady}, holds that relational structure is ontologically subsistent and prior to individual objects. It is ontologically fundamental in the sense of not supervening on the intrinsic properties of individuals. Quantum particles are contextually individuated: they are nodes in networks of relations. In so far as they are individuals, it is the relations among them that account for this.

This radical version of OSR runs counter to the widely held view that relations presuppose numerical diversity and thus cannot account for it. Esfeld and Lam (EL in what follows)~\cite{EsfeldLam,LamEsfeld} regard quantum mechanics as supporting a moderate version of OSR that does not waive the commitment to objects. While both versions of OSR deny that objects have an \emph{intrinsic} identity (constituted by intrinsic properties or a primitive thisness), the moderate version accepts numerical distinctness as primitive and denies that objects are reducible to relations. Instead, objects and relations imply each other: objects can neither exist nor be conceived without relations that hold between them, and relations can neither exist in the physical world nor be conceived as the structure of the physical world without objects that bear the relations.

The claim that structure is all there is leaves open how the structure is implemented, instantiated, or realized in the physical world. What is it that is structured? Ladyman and Ross \cite[p.~158]{LadyRoss} advocate a kind of neo-positivism according to which when questions like this arise it is time to stop: ``In our view, there is nothing more to be said about this that doesn't amount to empty words\dots. The `world-structure' just is and exists independently of us and we represent it mathematico-physically via our theories.'' Esfeld~\cite{Esfeld2013}, by contrast, insists that if OSR is to be a complete realism, it must specify how the structure in question is implemented. Otherwise it only provides a general scheme for an ontology of the physical domain, without spelling out a particular ontology, just as quantum mechanics does in the absence of a specific interpretation. Esfeld therefore suggests that OSR should be seen not as itself providing a complete ontology but as a possible guiding principle in the search for a quantum ontology. Apparently, though, its use as a guiding principle is limited, for, having examined some of the more popular interpretations, Esfeld concludes that none of them is in any obvious way compliant with OSR.

Esfeld's demand of completeness is something of a catch-22. If structure is ontologically subsistent and fundamental, OSR is incomplete in that it does not explain why the structure in question is physical and not just mathematical. And if a complete realism explains this, OSR cannot be a complete \emph{structural} realism since the physical realization of a given structure cannot be achieved by positing more structure, just as an interpretation of quantum mechanics cannot be achieved by writing down another formula.

The radical and the moderate ontic structural realist both appear unable or unwilling to recognize that the quantum world is structured in two distinct ways: there are (i)~the definite spatial relations between macroscopic objects, which are real \emph{per se}, and there are (ii)~the indefinite and reflexive spatial relations, which are instrumental in the manifestation of macroscopic objects, and which are described in terms of correlations between the possible outcomes of measurements. Thus EL~\cite{EsfeldLam} declare that the laws of physics ``describe relations among physical objects, and only relations, but without relations of measurement having a special status.'' 

Failing to recognize the difference between the two structures, along with the role that measurements play in rendering them distinct, the proponents of OSR fail to make further necessary distinctions, such as the distinction between relations and correlations. Thus Ladyman and Ross~\cite[p.~137]{LadyRoss} attribute to a pair of fermions in the singlet state ``the relation `is of opposite spin to'.'' This categorical statement is clearly false, for if the spins of the two particles are measured with respect to axes that are neither parallel nor antiparallel, they will not point in opposite directions. All that is warranted is conditional statements expressing (synchronic) correlations between possible measurement outcomes. 

EL~\cite{EsfeldLam}, for their part, having proposed that ``physical structures are networks of concrete, qualitative physical relations among objects that are nothing but what stands in these relations\dots,'' go on to state that ``the correlata are nothing but that what stands in the correlations,'' as if relations and correlations were interchangeable items. The correlata are \emph{not} nothing but what stands in the relations; they are events indicating the possession of a property (by a system) or a value (by an observable). The relata are indeed numerically distinct only on account of the relations that hold between them---their relative positions and orientations and the corresponding relative momenta---but intrinsically ``they'' are numerically identical and \emph{not} nothing but what stands in concrete, qualitative physical relations.

As far as quantum mechanics is concerned, OSR has been predicated on two features of the theory: the indiscernibility of particles of the same type and the failure of state separability in the case of entangled systems: whereas the quantum state of the whole system uniquely determines the states of its subsystems (i.e., their density operators), the latter do not uniquely determine the quantum state of the whole system. Both features are on display in EPR-Bohm experiments~\cite[pp. 611--622]{Bohm1951}, in which ``there is no fact of the matter which one of the two quantum objects prepared in the singlet state at the source of an EPR-Bohm experiment is later measured in the left wing and which one is measured in the right wing of the experiment''~\cite{Esfeld2013}. EL~\cite{LamEsfeld} interpret the failure of state separability as a violation of ``a cornerstone of atomism in the philosophy of nature,'' namely the principle of separability according to which the relations between objects (other than spatiotemporal relations) supervene on the intrinsic properties of objects:
\begin{quote}
instead of the intrinsic properties of the parts fixing the relations among them and thus the state of the whole, only the state of the whole fixes the relations among the parts, namely the superpositions of correlations that characterize entangled states.
\end{quote}
Whence EL~\cite{EsfeldLam} conclude that we ``cannot but take as fundamental the joint state of the whole, in the last resort the joint state of the whole world,'' and that ``the entangled states are the ways (modes) in which the quantum objects exist.'' 

These conclusions evince the lack of three necessary distinctions: the distinction between relations and correlations, the distinction between states qua sets of possessed properties and states qua probability algorithms, and the distinction between the supervenience of relations on properties and the supervenience of probability algorithms associated with composite systems on probability algorithms associated with component systems, which fails if the component systems are entangled. What is violated is not the principle that the relations between objects ought to supervene on the intrinsic properties of objects, but a principle which (to the best of my knowledge) has never been formulated, to wit, the principle that the probability algorithm associated with a composite system ought to supervene on the probability algorithms associated with its component systems. 

There are no ``superpositions of correlations.'' What leads to such awkward expressions is the failure to distinguish between relations (between objects) and correlations (between measurement outcomes). There are superpositions of quantum states associated with entangled systems, which superpositions are themselves quantum states, and which define correlations between the outcomes of measurements performed on the entangled systems. Instead of fixing relations among parts, the state of the whole determines correlations between outcomes of measurements to which the parts can be subjected. Nor can the joint state of the whole be taken as fundamental, for a probability algorithm presupposes the events to which it serves to assign probabilities. Entangled states are ways in which we describe composite systems that are \emph{prepared} in a certain ways, not ways in which they exist out of relation to what happens or is the case in the macroworld.

EL~\cite{EsfeldLam} characterize their version of OSR as a holism, which consists in ``regarding the whole world---or the domain of the world that one considers---as just one object in the last resort,'' where ``object'' has ``the same meaning as in atomism, namely `being that exists independently of other beings' (this is one sense of the traditional term `substance')'':
\begin{quote}
All the properties of that one object trivially are intrinsic properties, for there is nothing outside that object\dots. The idea is that there is an internal differentiation within the whole such that there are parts of the whole, and these parts have relational properties, that is, they stand in certain relations to one another.
\end{quote}
Since the central idea of the present paper also lends itself to being described as a holism (though I prefer to say what I mean without using this ambiguous term), I~am obliged to point out how it differs from the holism propounded by~EL~\cite{EsfeldLam}. 

To begin with, I fail to see how the latter can be cashed out consistently. I see no sense in assigning a quantum state to the whole universe, nor in regarding a quantum state as a description of a quantum system (except as a description in terms of correlations between measurement outcomes). To infer B from~A, where A is the non-supervenience of the probability algorithm associated with a whole on the probability algorithms associated with its parts, and B is the non-supervenience of the properties of the whole on the properties of its parts, is a \emph{non sequitur}. What is actually affirmed is not the existence of relations without (intrinsically distinct) relata but the existence of correlations without correlata, and there is no way for correlations between measurement outcomes to imply the existence of measurement outcomes.%
		\footnote{Mermin's thesis~\cite{Mermin1998}, according to which ``Correlations have physical reality; that which they correlate, does not,'' has a different import. Mermin did not claim that there are no correlata, only that they are not part of \emph{physical} reality. His idea (at the time) was that they belong to a larger reality which includes consciousness, and that the measurement problem only arises in this larger reality.}

The central idea of the present paper, according to which the macroworld is manifested through a differentiation of an intrinsically undifferentiated entity~$\cal E$, is not predicated on the entanglement of quantum systems. What exists independently of other beings is not the world as a whole but~$\cal E$, to whose differentiation the macroworld owes its existence. The differentiation consists in the coming into being of indefinite spatial relations. Insofar as this differentiation is internal, it is internal not to the whole world but to~$\cal E$, inasmuch as it is based on relations between $\cal E$ and itself. The self-differentiation of $\cal E$ through reflexive relations does not abrogate the numerical identity of the relata. It does, however, imply a new ontic dimension extending from $\cal E$ to the manifested world, a dimension across which the atemporal process of manifestation takes place. At one of its ends all fundamental particles in existence are numerically identical, at the other end they are numerically distinct.

The realism I propose goes beyond OSR in that it spells out how the relational structure of the physical world is realized. What gets structured is not a primitive, intrinsically unstructured multitude on which structure is imposed. This would be another edition of the Platonic-Aristotelian dualism of Matter and Form.%
		\footnote{For an insightful discussion of Plato's struggle to reconcile unity with multiplicity see Cornford~\cite{Cornford}. Twenty-four centuries later, we are still engaged in this struggle, although for the most part we are no longer aware of it.}
In fact, there is nothing that \emph{gets} structured. Structure comes to exist by an atemporal process of manifestation. A single, structureless entity manifests structure by entering into reflexive relations---both the indefinite relations which are instrumental in the manifestation of the macroworld, and the resulting definite relations which constitute the macroworld. It is true that without the relations there would be no relata, but there would still be that which has the power to enter into reflexive relations.


\begin{thebibliography}{99}
\bibitem{Zurek2003}
Zurek, W.H.: Decoherence, einselection, and the quantum origins of the classical. Rev. Mod. Phys. \textbf{75}, 715--775 (2003) 

\bibitem{Joosetal2003}
Joos, E., Zeh, H.D., Kiefer, C., Giulini, D.J.W., Kupsch, J., Stamatescu, I.-O.: Decoherence and the Appearance of a Classical World in Quantum Theory, 2nd Edition. Springer, New York (2003)

\bibitem{Schlosshauer07}
Schlosshauer, M.: Decoherence and the Quantum-to-Classical Transition. Springer, Berlin/Heidelberg (2007)

\bibitem{Falk}
Falkenburg, B.: Particle Metaphysics: A Critical Account of Subatomic Reality. Springer, Berlin Heidelberg (2007)

\bibitem{Mittelstaedt}
Mittelstaedt, P.: The Interpretation of Quantum Mechanics and the
Measurement Process. Cambridge University Press, Cambridge, MA (1998)

\bibitem{BLM}
Busch, P., Lahti, P.J., Mittelstaedt, P.: The Quantum Theory of Measurement, 2nd Revised Edition. Springer, Berlin (1996)

\bibitem{Hf74}
Hegerfeldt, G.: Remark on causality and particle localization. Phys. Ref. D \textbf{10}, 3320--3321 (1974)

\bibitem{Hf98}
Hegerfeldt, G.C.: Instantaneous spreading and Einstein causality in quantum theory. Ann. Phys. (Berlin) \textbf{7}, 716--25 (1998)

\bibitem{Hf2001}
Hegerfeldt, G.C.: Particle localization and the notion of Einstein causality. In A. Horzela, E. Kapuscik (Eds.), Extensions of Quantum Theory 3, 9--16. Apeiron, Montreal (2001)

\bibitem{Malament}
Malament, D.B.: In defense of dogma: Why there cannot be a relativistic quantum mechanics of (localizable) particles. In R. Clifton (Ed.), Perspectives on Quantum Reality, 1--10. Kluwer, Dordrecht (1996)

\bibitem{CH}
Clifton, R., Halvorson, H.: No place for particles in relativistic quantum theories. Philosophy of Science \textbf{69}, 1--28 (2002). Reprinted in J. Butterfield, H. Halvorson (Eds.), Quantum Entanglements: Selected Papers Rob Clifton, 225--261. Oxford University Press, Oxford (2004) 

\bibitem{LadyRoss}
Ladyman, J., Ross, D. (with Spurrett, D., Collier, J.): Every Thing Must Go: Metaphysics Naturalized. Oxford University Press, Oxford, New York (2007)

\bibitem{EsfeldLam}
Esfeld, M., Lam, V.: Structures as the objects of fundamental physics. In U. Feest, H.-J.  Rheinberger, G. Abel (Eds.), Epistemic Objects. Preprint \textbf{374}, 3--16. Max Planck Institute for the History of Science, Berlin (2009)

\bibitem{LamEsfeld}
Lam, V., Esfeld, M.: The structural metaphysics of quantum theory and general relativity. J.~Gen. Philos. Sci. \textbf{43}, 243--258 (2012)


\bibitem{Styeretal}
Styer, D.F., Balkin, M.S., Becker, K.M., Burns, M.R., Dudley, C.E., Forth, S.T., Gaumer, J.S., Kramer, M.A., Oertel, D.C., Park, L.H., Rinkoski, M.T., Smith, C.T., Wotherspoon, T.D.: Nine formulations of quantum mechanics. Am. J. P. \textbf{70}(3), 288--297 (2002) 

\bibitem{vonNeumann}
von Neumann, J.: Mathematische Grundlagen der Quantenmechanik (1932). English Translation: Mathematical Foundations of Quantum Mechanics. Princeton University Press, Princeton, NJ (1955)

\bibitem{Pauli1933}
Pauli, W.: Die allgemeinen Prinzipien der Wellenmechanik (1933). English Translation: General Principles of Quantum Mechanics. Springer, Berlin (1980)

\bibitem{Dirac}
Dirac, P.A.M.: The Principles of Quantum Mechanics. Clarendon Press, Oxford (1958)


\bibitem{FHS}
Feynman, R.P., Hibbs, A.R., Styer, D.F.: Quantum Mechanics and Path Integrals, Emended Edition. Dover Publications, Mineola, NY (2005)

\bibitem{Mohrhoff-QMexplained}
Mohrhoff, U.: Quantum mechanics explained. Int. J. Q. Inf. \textbf{7}(1), 435--458 (2009)

\bibitem{SEW1991}
Scully, M.O., Englert, B.-G., Walther, H.: Quantum optical tests of complementarity. Nature \textbf{351}(6322), 111--116 (1991)

\bibitem{ESW1994}
Englert, B.-G., Scully, M.O., Walther, H.: The duality in matter and light. Sci. Am. (Int. Ed.) \textbf{271}(6), 56--61 (December 1994)

\bibitem{Mohrhoff-ESW}
Mohrhoff, U.: Objectivity, retrocausation, and the experiment of Englert, Scully, and Walther. Am. J. Phys. \textbf{67}(4), 330--335 (1999)

\bibitem{Fuchs}
Fuchs, C.: In M. Schlosshauer (Ed.), Elegance and Enigma: The Quantum Interviews. Springer, Berlin Heidelberg (2011)

\bibitem{vG}
von Glasersfeld, E.: An Introduction to Radical Constructivism. In P. Watzlawick (Ed.), The Invented Reality, 17--40. Norton, New York (1984)


\bibitem{Feynman65}
Feynman, R.P., Leighton, R.B., Sands, M.: The Feynman Lectures in Physics, Vol.\ 3. Addison-Wesley, Boston, MA (1965)

\bibitem{Tonomura89}
Tonomura, A., Endo, J., Matsuda, T., Kawasaki, T.: Demonstration of single-electron buildup of an interference pattern. Am. J. P. \textbf{57}(2), 117--120 (1989) 

\bibitem{Arndtetal}
Arndt, M., Nairz, O., Vos-Andreae, J., Keller, C., van der Zouw, G., Zeilinger, A.: Wave-particle duality of C$_{60}$ molecules. Nature \textbf{401}, 680--682 (1999)

\bibitem{Bohr34}
Bohr, N.: Atomic Theory and the Description of Nature. Cambridge University Press, Cambridge (1934) 

\bibitem{Bohr49}
Bohr, N.: Discussions with Einstein on epistemological problems in atomic physics. In P.A. Schilpp (Ed.), Albert Einstein: Philosopher--Scientist, 201--241. Library of Living Philosophers, Evanston, IL (1949)

\bibitem{Bohr63}
Bohr, N.: Quantum physics and philosophy---causality and complementarity. In N. Bohr, Essays 1958--1962 on Atomic Physics and Human Knowledge, 1--7. Interscience Publishers (John Wiley \& Sons), New York/London (1963)

\bibitem{Hubel95}
Hubel, D.H.: Eye, Brain, and Vision. Scientific American Library, New York (1995)

\bibitem{Enns04}
Enns, J.T.: The Thinking Eye, the Seeing Brain: Explorations in Visual Cognition. Norton \& Company, New York/London (2004)

\bibitem{Clark}
Clark, A.: A Theory of Sentience. Oxford University Press, New York (2000)

\bibitem{vW1980}
von Weizs\"acker, C.F.: The Unity of Nature. Farrar, Straus, and Giroux, New York (1980)

\bibitem{Hilgevoord98}
Hilgevoord, J.: The uncertainty principle for energy and time II. Am. J. P. \textbf{66}(5), 396--402 (1998)


\bibitem{Bell1990}
Bell, J.S.: Against ``measurement.'' Physics World, 33--40 (August 1990)

\bibitem{BubBW}
Bub, J.: Bananaworld: Quantum mechanics for primates. arXiv:1211.3062v2 [quant-ph] (2013)

\bibitem{BubClifton}
Bub, J., Clifton, R.: A uniqueness theorem for ``no collapse'' interpretations of quantum mechanics. Stud. Hist. Phil. Mod. Phys. \textbf{27}, 181--219 (1996)

\bibitem{MohrhoffBW}
Mohrhoff, U.: Interpreting bananaworld: a response to Bub's quantum mechanics for primates. arXiv:1212.3606v1 [quant-ph] (2012)

\bibitem{BellCH}
Bell, J.S.: On wave-packet reduction in the Coleman-Hepp model.
In J.S. Bell, Speakable and Unspeakable in Quantum Mechanics, 45--51. Cambridge University Press, Cambridge (1987)

\bibitem{Hepp}
Hepp, K.: Quantum theory of measurement and macroscopic observables. Helv. Phys. Acta \textbf{45}, 237--248 (1972) 

\bibitem{BGL}
Busch, P., Grabowski, M., Lahti, P.: Operational Quantum Physics. Springer, Berlin (1995)

\bibitem{Bushi}
Busch, P., Shimony, A.: Insolubility of the quantum measurement problem for unsharp observables. Stud. Hist. Phil. Mod. Phys. \textbf{27}, 397--404 (1996)

\bibitem{Bush96}
Busch, P.: Can ``unsharp objectification'' solve the quantum measurement problem? Int. J. Theor. Phys. \textbf{37}, 241--247 (1998)


\bibitem{SEP-French}
French, S.: Identity and Individuality in Quantum Theory. In E.N. Zalta (Ed.), The Stanford Encyclopedia of Philosophy. URL = http://plato.stanford.edu/archives/sum2011/entries/\break qt-idind (Summer 2011 Edition) 

\bibitem{FrenchKrause}
French, S., Krause, D.: Identity in Physics: A Historical, Philosophical, and Formal Analysis. Oxford University Press, Oxford (2006)

\bibitem{Esfeld2013}
Esfeld, M.: Ontic structural realism and the interpretation of quantum mechanics. Eur. J. Philos. Sci. \textbf{3}(1), 19--32 (2013)


\bibitem{UlfBohr}
Ulfbeck, O., Bohr, A.: Genuine fortuitousness. Where did that click come from? Found. Phys. \textbf{31}(5), 757--774 (2001)

\bibitem{Mohrhoff-clicks}
Mohrhoff, U.: Making sense of a world of clicks. Found. Phys. \textbf{32}(8), 1295--1311 (2002)

\bibitem{GS}
Grupen, C., Shwartz, B.: Particle Detectors, Second Edition. Cambridge Unversity Press, Cambridge (2008)


\bibitem{Heisenberg}
Heisenberg, W.: The Physical Principles of Quantum Theory. English translation by C. Eckart and F.C. Hoyt. University of Chicago Press, Chicago, IL (1930)


\bibitem{Hoffman98}
Hoffman, D.D.: Visual Intelligence: How We Create What We See. Norton \& Company, New York (1998)

\bibitem{EPPP}
Elementary-Particle Physics Panel, National Research Council (U.S.): Elementary-Particle Physics. National Academy Press, Washington, DC (1998)

\bibitem{Nicolai}
Nicolai, H.: Quantum gravity: the view from particle physics. Invited Lecture at the conference Relativity and Gravitation: 100 Years after Einstein in Prague, June 25--29, 2012, Prague, Czech Republic. arXiv:1301.5481 [gr-qc] (2013)


\bibitem{LL77}
Landau, L.D., Lifshitz, E.M.: Quantum Mechanics, 3rd Edition. Pergamon, Oxford (1977)

\bibitem{Redhead1990}
Redhead, M.: Explanation. In D. Knowles (Ed.), Explanation and its Limits, 135--154. Cambridge University Press, Cambridge, U.K. (1990)

\bibitem{Sellars}
Sellars, W.: Philosophy and the scientific image of man. In: R. Colodny (Ed.), Frontiers of Science and Philosophy, 35--78. University of Pittsburgh Press, Pittsburgh, PA (1962)


\bibitem{Eddington}
Eddington, A.S.: The Nature of the Physical World. MacMillan, New York (1929)

\bibitem{Maudlin}
Maudlin, T.: Descrying the world in the wave function. The Monist \textbf{80}(1), 3--23 (1997)

\bibitem{Phillips}
Phillips, S.: Classical Indian Metaphysics. Open Court, Chicago/La Salle, IL (1995)

\bibitem{Aurobindo}
Sri Aurobindo: The Life Divine. Sri Aurobindo Ashram Publication Department, Pondicherry (2005)

\bibitem{Mohrhoff-CQM}
Mohrhoff, U.: Consciousness in the quantum world: An Indian perspective. In: A. Corradini and U. Meixner (Eds.), Quantum Physics Meets the Philosophy of Mind: New Essays on the Mind-Body Relation in Quantum-Theoretical Perspective. To be published by de Gruyter, Berlin/New York (2014)

\bibitem{SchrLife}
Schr\"odinger, E.: What is Life? The Physical Aspect of the Living Cell. Cambridge University Press, Cambridge (1944). Reproduced in K. Wilber (Ed.), Quantum Questions: Mystical Writings of the World's Greatest Physicists, 92--95. Shambala, Boston (2001)

\bibitem{Russell}
Russell, B.: The Analysis of Matter. Routledge, London (1927/1992)

\bibitem{BarrowTipler}
Barrow, J.-D., Tipler, F.J.: The Anthropic Cosmological Principle. Oxford University Press, Oxford (1988)

\bibitem{Rees}
Rees, M.: Just Six Numbers: The Deep Forces That Shape the Universe. Basic Books, New York (2000)

\bibitem{Bostrom}
Bostrom, N.: Anthropic Bias: Observation Selection Effects in Science and Philosophy. Routledge, New York (2002)

\bibitem{Mohrhoff-book}
Mohrhoff, U.: The World According to Quantum Mechanics. World Scientific, Singapore (2011) 

\bibitem{Mohrhoff-justso}
Mohrhoff, U.: Why the laws of physics are just so. Found. Phys. \textbf{32}(8), 1313--1324 (2002)

\bibitem{Wilczek}
Wilczek, F.: The Lightness of Being: Mass, Ether, and the Unification of Forces. Basic Books, New York (2008)


\bibitem{SEP-Ladyman}
Ladyman, J.: Structural Realism. In E.N. Zalta (Ed.), The Stanford Encyclopedia of Philosophy. URL = http://plato.stanford.edu/archives/sum2013/entries/structural-realism (Summer 2013 Edition)

\bibitem{FrenchLady}
French, S., Ladyman, J.: Remodelling structural realism: Quantum physics and the metaphysics of structure. Synthese \textbf{136}, 31--56 (2003)

\bibitem{Bohm1951}
Bohm, D.: Quantum theory. Prentice-Hall, Englewood Cliffs, NJ (1951)

\bibitem{Mermin1998}
Mermin, N.D.: What is quantum mechanics trying to tell us? Am. J. Phys. \textbf{66}, 753--767 (1998)

\bibitem{Cornford}
Cornford, F.M.: Plato and Parmenides. Kegan Paul, Trench, Trubner \& Co., London (1939)

\end{thebibliography}
\end{document}